\documentclass[aps,prb,reprint,superscriptaddress,showpacs]{revtex4-1}
\usepackage{color}
\usepackage{graphicx}
\usepackage[american]{babel}
\usepackage[utf8]{inputenc}
\usepackage[T1]{fontenc}
\usepackage{bbm,ae,aecompl}
\usepackage{amsmath}
\usepackage{amssymb}
\usepackage{amstext}
\usepackage{amsthm}
\usepackage{latexsym}
\usepackage{verbatim}
\usepackage{natbib}

\usepackage{ulem}   % For strikethrough

\begin{document}

\title{Impact of the 2 Fe unit cell on the electronic structure \\ measured by ARPES in iron pnictides.}

\date{\today}

\author{V. Brouet}
%\affiliation{Laboratoire de Physique des Solides, Universit\'{e} Paris-Sud, UMR 8502, B\^at. 510, 91405 Orsay, France}

\author{M. Fuglsang Jensen}
%\email{jensen@lps.u-psud.fr}
%\affiliation{Laboratoire de Physique des Solides, Universit\'{e} Paris-Sud, UMR 8502, B\^at. 510, 91405 Orsay, France}

\author{Ping-Hui Lin}
%\email{jensen@lps.u-psud.fr}
\affiliation{Laboratoire de Physique des Solides, Universit\'{e} Paris-Sud, UMR 8502, B\^at. 510, 91405 Orsay, France}

%\author{A. Nicolaou}
%\affiliation{Synchrotron SOLEIL, L'Orme des Merisiers, Saint-Aubin-BP 48, 91192 Gif sur Yvette, France}

\author{A. Taleb-Ibrahimi}
%\affiliation{Synchrotron SOLEIL, L'Orme des Merisiers, Saint-Aubin-BP 48, 91192 Gif sur Yvette, France}

\author{P. Le F\`evre}
%\affiliation{Synchrotron SOLEIL, L'Orme des Merisiers, Saint-Aubin-BP 48, 91192 Gif sur Yvette, France}

\author{F. Bertran}
\affiliation{Synchrotron SOLEIL, L'Orme des Merisiers, Saint-Aubin-BP 48, 91192 Gif sur Yvette, France}

\author{Chia-Hui Lin}
%\affiliation{Condensed Matter Physics and Materials Science Department,
%Brookhaven National Laboratory, Upton, New York 11973, USA}
%\affiliation{Department of Physics and Astronomy, Stony Brook University, Stony Brook, New York 11794, USA}

\author{Wei Ku}
\affiliation{Condensed Matter Physics and Materials Science Department,
Brookhaven National Laboratory, Upton, New York 11973, USA}
\affiliation{Department of Physics and Astronomy, Stony Brook University, Stony Brook, New York 11794, USA}

\author{A. Forget}
%\affiliation{Service de Physique de l'Etat Condens\'{e}, Orme des Merisiers, CEA Saclay (CNRS URA 2464), 91191 Gif sur Yvette Cedex, France}

\author{D. Colson}
\affiliation{Service de Physique de l'Etat Condens\'{e}, Orme des Merisiers, CEA Saclay, CNRS-URA 2464, 91191 Gif sur Yvette Cedex, France}

\begin{abstract}
In all iron pnictides, the positions of the ligand alternatively above and below the Fe plane create 2 inequivalent Fe sites. This results in 10 Fe $3d$ bands in the electronic structure. However, they do not all have the same status for an ARPES experiment. There are interference effects between the 2 Fe that modulate strongly the intensity of the bands and that can even switch their parity. We give a simple description of these effects, notably showing that ARPES polarization selection rules in these systems cannot be applied by reference to a single Fe ion. We show that ARPES data for the electron pockets in Ba(Fe$_{0.92}$Co$_{0.08}$)$_2$As$_2$ are in excellent agreement with this model. We observe both the total suppression of some bands and the parity switching of some other bands. Once these effects are properly taken into account, the structure of the electron pockets, as measured by ARPES, becomes very clear and simple.  By combining ARPES measurements in different experimental configurations, we clearly isolate {\it each} band forming {\it one} of the electron pockets. We identify a deep electron band along one ellipse axis with the $d_{xy}$ orbital and a shallow electron band along the perpendicular axis with the $d_{xz}$/$d_{yz}$ orbitals, in good agreement with band structure calculations. We show that the electron pockets are warped as a function of $k_z$ as expected theoretically, but that they are much smaller than predicted by the calculation.

\end{abstract}

\maketitle

\section{Introduction}
There is a consensus that the multiband nature of iron pnictides is essential for defining their electronic properties. Many models for the superconducting and antiferromagnetic orders heavily rely on the interaction between different electron and hole Fermi Surface (FS) sheets \cite{MazinPRL08,Paglione:2010p271}. However, despite many ARPES investigations of iron pnictides, the structure of the electron pockets are still poorly understood. They appear more difficult to clearly detect in ARPES than the hole pockets. Usually, only part of the expected ellipse is observed \cite{ThirupathaiahPRB10,MalaebJPSJ09,LiuNatPhys10}, leaving unclear what is really the size of the pocket or whether the electronic properties are isotropic around the ellipse. Recently, an ARPES study resolved them in more details \cite{ZhangPRB11}, but proposed a structure so different from band structure calculations, that it clearly calls for more investigation. On the contrary, other experimental techniques, such as transport \cite{RullierAlbenquePRL09,FangPRB09}, Raman \cite{ChauvierePRB10,MuschlerPRB09} or quantum oscillations experiments \cite{ColdeaPRL08,ShishidoPRL10}, seem to detect predominantly electrons, as if they had longer lifetimes. It would be desirable to measure the Fermi velocities and lifetimes of each hole and electron bands separately by ARPES, to understand the origin of this difference. This has not been possible so far, due to the low resolution obtained on electron pockets.
%===============================
\begin{figure*}[tbp]
\centering
\includegraphics[width=0.9\textwidth]{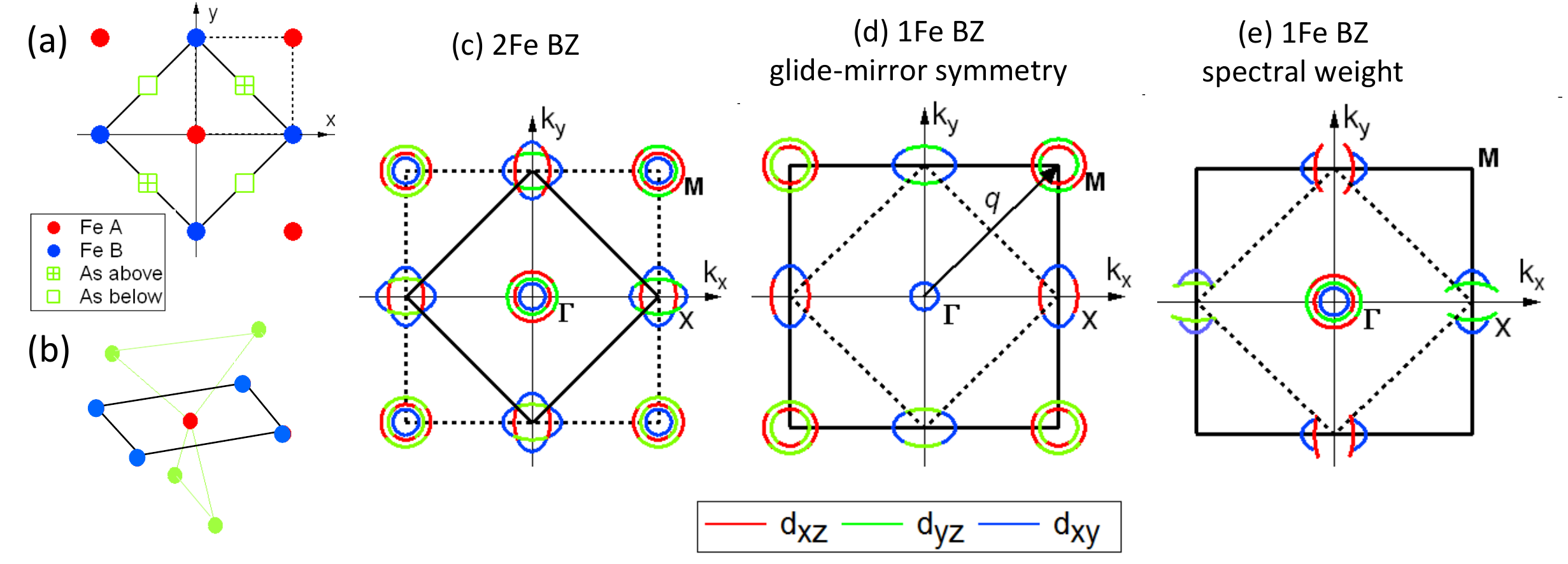}
\caption{(a) Structure of one FeAs slab. The thick black line represents the 2D unit cell with inequivalent Fe at the center (A) and the corners (B), because of the position of the neighboring As. The dotted black line defines the basic Fe square. (b) Tetrahedral As environment of one Fe. (c) 2D Fermi Surface expected in the 2Fe BZ built on the unit cell. The color code for different orbital characters. The pockets at $\Gamma$ and M are hole pockets, those at X are electron pockets. The dotted square is the 1Fe BZ built on the basic Fe square. (d) FS in the 1Fe BZ. The additional pockets in the 2Fe BZ can be obtained by folding with respect to the 2Fe BZ boundaries or, equivalently,  by translation by $q_F$. We follow here ref.\cite{Andersen2011} for the definition of main and folded pockets, other authors reverse the $\Gamma$ and M points \cite{Graser2010}. (e) Sketch of the expected spectral distribution, following the analysis of section II.}
\label{sketches}
\end{figure*}
%===============================

In this paper, we present a complete investigation of the electron pockets in Ba(Fe$_{0.92}$Co$_{0.08}$)$_2$As$_2$. This compound corresponds to optimal Co doping for superconductivity ($T_c$=23K), but we have obtained similar spectra for BaFe$_2$As$_2$ \cite{JensenPRB11} and other Co dopings. A complexity of ARPES in iron pnictides is that there are 2 Fe per unit cell, which doubles the number of bands. In section II, we give a simple description of the folding from the 1Fe to 2Fe Brillouin Zones (BZ), which yields a qualitative understanding of the spectral weight distribution on the main and folded bands. To our knowledge, folding effects were only discussed before in ARPES in the context of a {\it weak} potential \cite{VoitScience2000,Brouet:2008}. This case is different, because the As potential is strong, but essentially only its symmetry changes between the 2 inequivalent Fe sites. We show that this does modulate strongly the spectral weight, but in a way that depends on the orbital symmetry. Understanding this allows to identify the bands that should be observed in ARPES. This also explains that the parities of the bands depends on the relative phase between the 2 Fe of the unit cell. This implies that the polarization selection rules for ARPES are considerably different from those generally used in this community (see section II-C) and some bands appear as if they have \lq\lq{}switched\rq\rq{} parity. In section III, we present our ARPES data, where one full electron ellipse is clearly resolved. We first compare our measurements with band structure calculations, which allows to identify the band characters. Using the correct polarization selection rules solves the problems encountered in ref. \cite{ZhangPRB11} and reconciles ARPES with band calculations. We then compare in section III-B the measurements with simulations using the unfolded band structure of ref.\cite{LinWeiKu}. This gives the spectral weights of the main and folded bands, which are found in excellent agreement with our measurement. We finally study in III-C the $k_z$ dependence of the different electron bands. The size of the pockets and their $k_z$ dependence are smaller than predicted in theory, confirming a tendency previously noted \cite{ColdeaPRL08,OrtenziPRL09,BrouetPRB09}. A shift of the calculated electron bands as large as 150meV is needed to fit the experiment.

$\\$

Single crystals were grown using a FeAs self-flux method and were studied in details by transport measurements \cite{RullierAlbenquePRL09}.  ARPES experiments were carried out at the CASSIOPEE beamline at the SOLEIL synchrotron, with a Scienta R4000 analyser, an angular resolution of $0.2^{\circ}$ and an energy resolution better than $15$ meV. Band structure calculations were perfomed within the local density approximation, using the Wien2K package \cite{Wien2k}, with the experimental structure of BaFe$_2$As$_2$ and a doping of 10\% treated in the virtual crystal approximation.

\section{Bands in the 1 Fe and 2 Fe BZ}

In all iron pnictides, there are 2 inequivalent Fe per unit cell, as sketched in red and blue in Fig. 1(a), and called sites A and B in the following. The inequivalency is due to the position of the surrounding As, which always form a tetrahedra, but with the edge oriented along 110 either above the Fe plane, as in Fig. 1(b), or below. The As potential created at the 2 inequivalent Fe sites is equal, as first approximation, but with an opposite symmetry with respect to z.

In the 2Fe BZ corresponding to this structural unit cell [thick line in Fig. 1(c)], there are 10 Fe $3d$ bands to consider, making the band structure quite complicated to decipher. To simplify this problem, many theoreticians use a 1Fe BZ, as in Fig. 1(d), which is based on the small Fe square [dotted line in Fig. 1(a)], and where the number of bands is divided by two. It is well known that the other bands can be obtained to a good approximation by folding with respect to the 2Fe BZ boundaries \cite{MazinPRL08,Graser2010,Andersen2011}. Curiously, very little attention has been paid to this underlying structure by ARPES authors. It is usually considered as a simple convention to work in the 1Fe or 2Fe BZ. On the contrary, the distribution of spectral weight on the main (1Fe) and folded bands is an intrinsic feature of the electronic structure. We will show that this modulates strongly the ARPES intensities of the bands and that it is very useful to understand how this works to correctly interpret ARPES spectra.

In the following, we choose the $x$ and $y$ axis in the Fe-Fe direction, as in Fig. 1(a), the origin at one Fe site, and we call $a$ the Fe-Fe distance. The orbitals are also defined with respect to these axes, which are those of the 1Fe BZ. In Fig. 1, we consider only 2 dimensional (2D) BZ. In reality, there is a component of the folding along $k_z$ for BaFe$_2$As$_2$, because it is body centered \cite{Graser2010}. This will be considered later in section III. We label the high symmetry points X and M according to the 2Fe BZ, to keep the same notations between sections II and III.

\subsection{Folding bands from the 1Fe to 2Fe BZ}
	
	We first consider a square lattice of Fe. We define Bloch functions $\psi^{1Fe}_{k,\chi}(r)$ built from atomic $\chi$(r) orbitals.
\begin{equation}
\psi^{1Fe}_{k,\chi}(r)=\sum_{R}e^{ik \cdot R}\chi(r-R)
  \label{equa1}
\end{equation}
$\chi$(r) is one of the $3d$ Fe orbitals and the sum runs over all Fe sites (for clarity, we will omit the index $\chi$ in the following). At $\Gamma$, all orbitals are in phase and a spatial representation of $\psi^{1Fe}_k(r)$ for $d_{xy}$ is given in Fig. 2(b). At X, i.e. k=($\pi$/a,0), there is an additional dephasing $\phi=k \cdot R$ due to the Bloch term. For the Fe sites along $y$, $\phi=0$ and for the other sites at ($\pm$a,0), $\phi=\pi$. This is sketched in Fig. 2(c), again for $d_{xy}$.

The BZ of this square lattice is the 1Fe BZ. However, one could decide to work in the 2Fe BZ and obtain an equivalent description of these bands by folding $\psi^{1Fe}$ with respect to the new zone boundaries, or, equivalently, translating them by $q=(\pi/a,\pi/a)$ [see Fig. 1(d)]. This creates a new band $\psi_{k}^{Folded}(r)$.
\[\psi_{k}^{Folded}(r)=\psi_{k+q}^{1Fe}(r)=\sum_{R}e^{iq \cdot R}e^{ik \cdot R}\chi(r-R)\]

For the Fe on sites A, $R=(0,0)$, yielding $e^{iq \cdot R}=1$. For sites B, $R=($$\pm$$a,0)$ or $(0,$$\pm$$a)$, yielding $e^{iq \cdot R}=-1$. Hence,
\[\psi_k^{Folded}(r)=\sum_{R_A}e^{ik \cdot R_A}\chi(r-R_A)-\sum_{R_B}e^{ik \cdot R_B}\chi(r-R_B)\]
In other words, the Fe on sites B are dephased by $\pi$ compared to sites A. This is represented in Fig. 2(d) for $d_{xy}$ at $\Gamma$. An equivalent description of the 2 bands in the 2Fe BZ is then to define a linear combination of the 2 Fe with sites A and B in-phase or out-of-phase. We will call $\psi$ the in-phase bands and $\psi^{*}$ the out-of-phase bands, like $d_{xy}$ in Fig. 2(b) and $d_{xy}$* in Fig. 2(d). %\[\psi_{k}(r)=\sum_{R_A}e^{ik.R_A}\chi(r-R_A)+\sum_{R_B}e^{ik.R_B}\chi(r-R_B)\]
%\[\psi^{*}_{k}(r)=\sum_{R_A}e^{ik.R_A}\chi(r-R_A)-\sum_{R_B}e^{ik.R_B}\chi(r-R_B)\]

%===============================
\begin{figure}[tbp]
\centering
\includegraphics[width=0.45\textwidth]{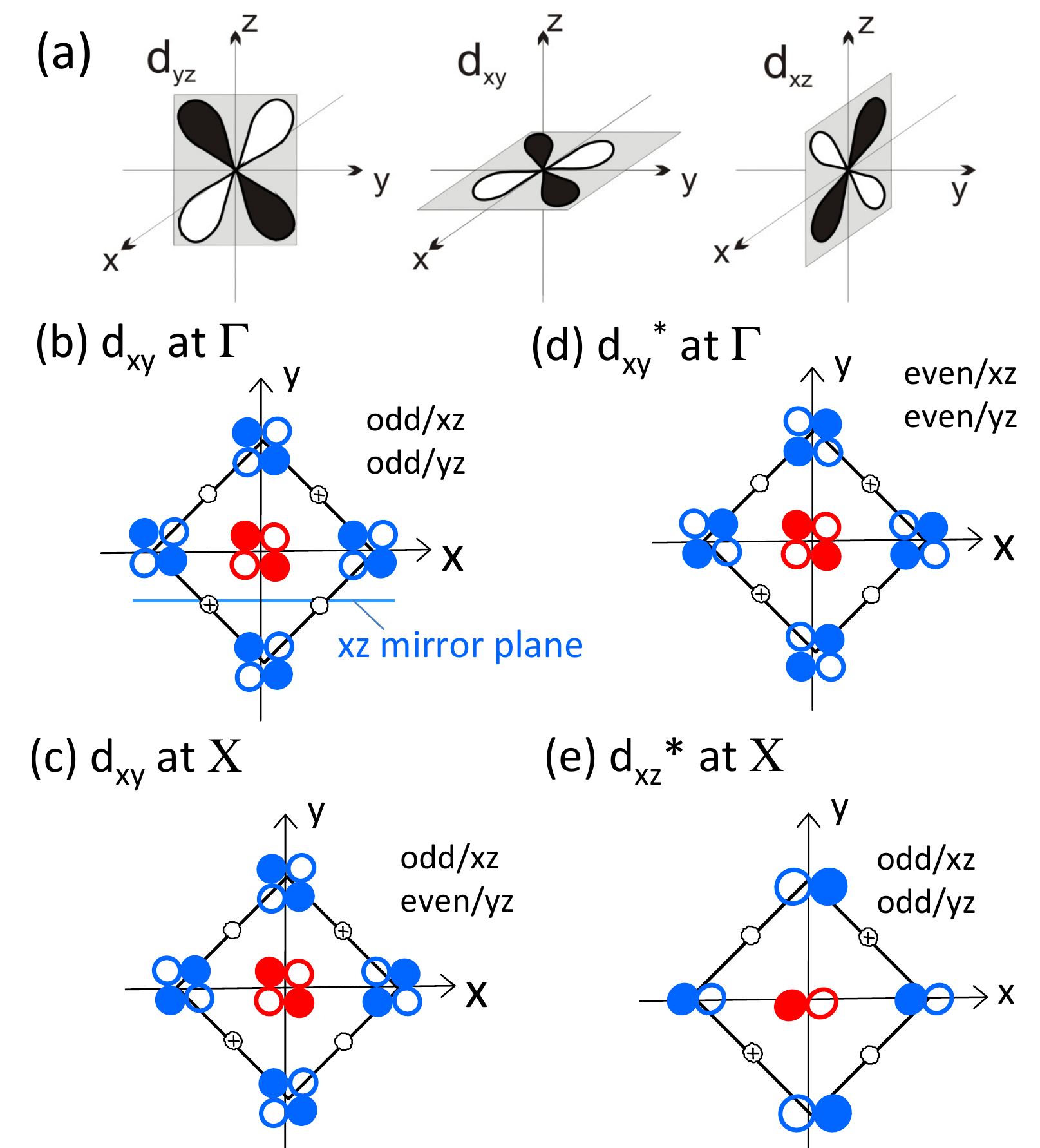}
\caption{(a) Sketch of the three t$_{2g}$ orbitals, $d_{xy}$, $d_{xz}$ and $d_{yz}$. (b-e) Relative orientations of orbitals in real space for the following orbitals and points of the reciprocal space. (b) $d_{xy}$ at $\Gamma$, (c) $d_{xy}$* at X($\pi$/a,0), (d) $d_{xy}$* at $\Gamma$, (e) $d_{xz}$* at X. The blue line in (b) indicates the $xz$ mirror plane.}
\label{Fig2}
\end{figure}
%===============================

The choice of the 1Fe or 2Fe BZ is of course dictated by physical considerations. Usually, the 2Fe BZ is required by the presence of a small potential V making the 2 Fe in the unit cell inequivalent. The eigenstates $\psi^{I}$ and $\psi^{II}$ will be a linear combination of $\psi^{1Fe}$ and $\psi^{Folded}$ (or equivalently of $\psi$ and $\psi^{*}$). The coupling will for example open gaps at the bands crossings, proportional to the folding potential V. In iron pnictides, the real bands may also hybridize with other orbitals having the same symmetry and acquire some finite weight from different orbital character. Nevertheless, if these perturbations remain sufficiently small, the eigenstates behave essentially as the bands we have defined here, and they keep the same symmetry, so that these bands are very useful guides to simply visualize the situation.

\subsection{Spectral weight distribution }

Since the final state of the photoelectron of the ARPES process corresponds to a free-electron like state far away from the sample, the ARPES spectral weight is naturally given by the 1Fe picture with similar translational symmetry.The spectral weights $w^{I,II}_k$ of the 2 eigenstates $\psi^{I,II}$ can be obtained by the projection on the basis of the 1Fe BZ. With the previous notations,
\begin{equation}
w^{I,II}_k=|<\psi^{1Fe}_k|\psi^{I,II}_k>|^2
  \label{equa2}
\end{equation}
For a weak folding potential, the mixing between $\psi^{1Fe}$ and $\psi^{Folded}$ is rather small, except near band crossings, and the spectral weight remains concentrated along the bands of the 1Fe BZ.  Many such examples can be found in ARPES literature \cite{VoitScience2000,Brouet:2008}. In iron pnictides, the potential due to the As is certainly not weak, as the As is essential to define the overlap between the Fe \cite{Lee:2009}. On the other hand, this potential is very similar at each Fe site, in fact, it is essentially reversed with respect to z. This leads to original folding effects, where half the Fe bands do have weaker spectral weight, but not those defined as the 1Fe bands in Fig. 1(d).

To perform the calculation of Fig. 1(d), all Fe must be equivalent. As the As potential cannot be neglected, a way to achieve this is to introduce a virtual operation symmetry, which changes the sign of z when going from Fe A to B. This is the glide-mirror symmetry described by Andersen and Boeri\cite{Andersen2011} under which all Fe become equivalent. As $d_{xy}$ is symmetric with respect to z [see Fig. 2(a)], it does not change between the two sites under this glide-mirror symmetry. On the contrary, $d_{xz}$ and $d_{yz}$ change sign. {\it It follows that the \rq\rq{}main bands\rq\rq{} of this 1Fe BZ are the one we defined as in-phase for $d_{xy}$, but out-of-phase for $d_{xz}$ and $d_{yz}$}. Eq. (2) predicts that the spectral weight will be close to 1 for all in-phase bands and to 0 for all out-of-phase bands. This will give rise to incomplete ellipses and inequivalency between $\Gamma$ and M, as sketched in Fig. 1(e). This is also a 1Fe representation of the electronic structure, but where the real space coordinate has not been redefined by utilizing glide-mirror symmetry. Instead, a flat reciprocal space of 1Fe BZ has been constructed by treating the translational symmetry breaking as the perturbation. This definition retains the symmetry of the physical space where ARPES works.

%If one does not re-define the real space coordinate by utilizing glide-mirror symmetry, a flat reciprocal space of 1Fe BZ can be constructed for $\psi^{1Fe}$ as in Eq. (1) by treating the translational symmetry breaking as the perturbation in this representation. This definition also helps to decipher the structure of band folding effect and retains the symmetry of the physical space where ARPES works. Thus,

%===================================================
%\begin{table}[b]
%\caption{\label{''Table 1''} Symmetries with respect to mirror planes $xz$ or $yz$ of the different bands forming the electron pocket along $k_x$. The star refers to out-of-phase orbitals, as defined in section II-A. The directions $\Gamma$X and XM are defined in the 1Fe BZ of Fig. 1. They are equivalent to ZX and X$\Gamma$ in the 2Fe BZ of BaFe$_2$As$_2$ used in Fig. \ \ref{FSfig}.}
%\begin{tabular}{|c|c|c|c|c|}
%\hline
 %     & $d_{xz}^{*}$ & $d_{yz}$ & $d_{xy}$ & $d_{xy}^{*}$ \\
 % \hline
  %   $\Gamma$X &odd/xz&odd/xz&odd/xz&even/xz\\
 %   XM  &odd/yz&odd/yz&even/yz&odd/yz\\
 %  \hline
%\end{tabular}
%\end{table}
%===================================================

%===============================
\begin{figure}[tbp]
\centering
\includegraphics[width=0.45\textwidth]{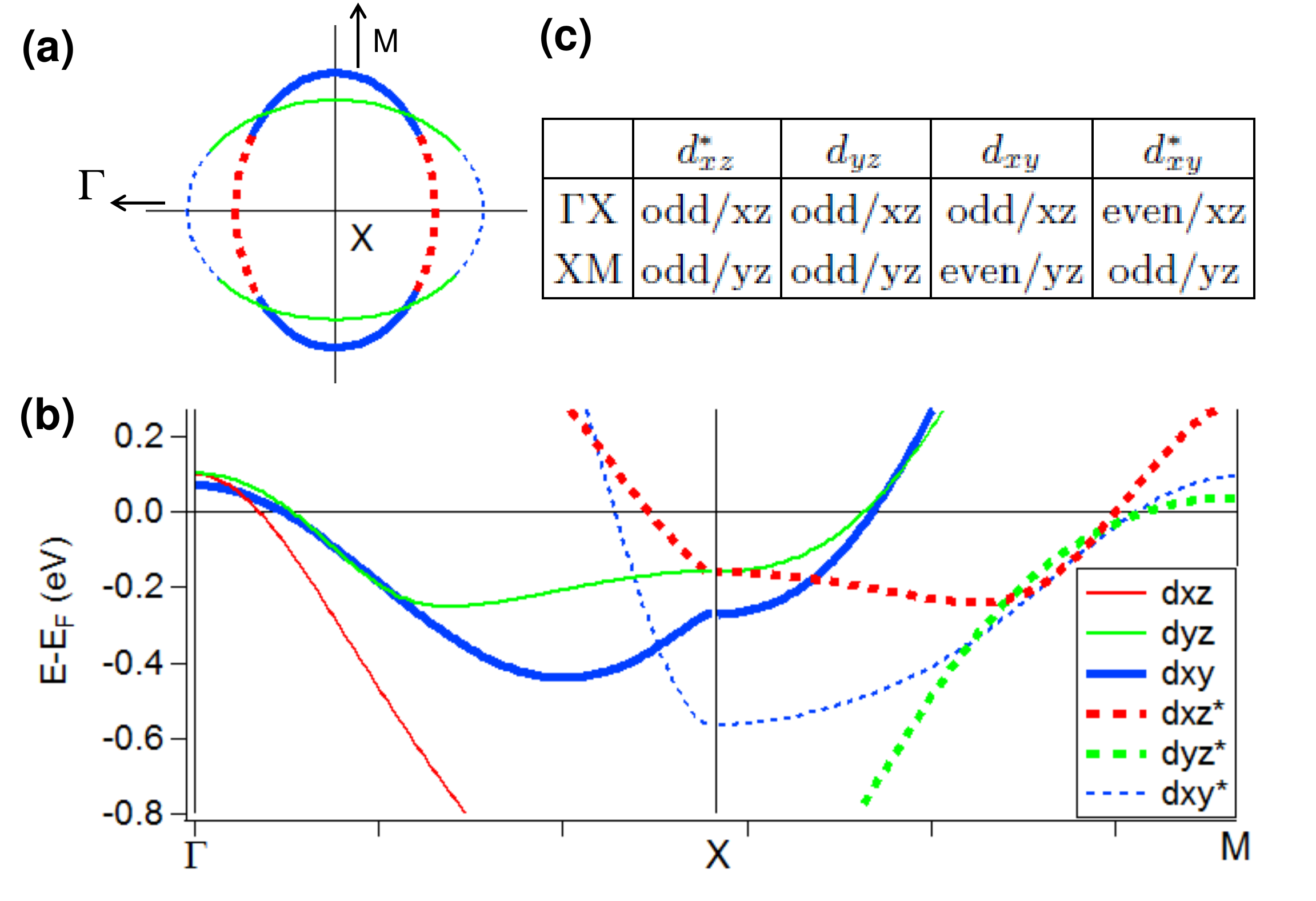}
\caption{(a) Sketch of the bands forming the electron pockets. Thick and thin lines indicate main and folded bands in the sense of the 1Fe BZ with glide-mirror symmetry. Solid and dotted lines represent the in-phase and out-of-phase orbitals. Colors indicate the main orbital character. (b) Sketch of the dispersion of the different bands along the path $\Gamma$XM. The color sketches the main orbital character, although there may be hybridization with other orbitals. (c) Symmetries of the different bands forming the electron pocket at X.
%The star refers to out-of-phase orbitals, as defined in section II-A. The directions $\Gamma$X and XM are defined in the 1Fe BZ of Fig. 1. They are equivalent to ZX and X$\Gamma$ in the 2Fe BZ of BaFe$_2$As$_2$ used in Fig. \ \ref{FSfig}.
}
\label{last}
\end{figure}
%===============================

The recently proposed band structure ``unfolding'' method~\cite{WeiKuPRL2010} provides a simple mean to calculate such an unfolded band structure from first-principles calculation. By representing the one-particle spectral function in the 1Fe Bloch basis, the above phase interference effect (associated with the Fe positions) that was hidden in the 2Fe Bloch functions now enters explicitly the theoretical spectral weight. This was recently worked out explicitly in ref. ~\cite{LinWeiKu}.
%The calculation of spectral weight was recently worked out explicitly in ref.  \cite{LinWeiKu}. In this work, Green\rq{}s functions are transformed from 2Fe BZ basis into 1Fe BZ basis by utilizing the symmetry respecting Wannier orbitals. The calculated spectral functions absorb the structure factor and the in-phase/out-of-phase effects.
Essentially, it finds that the spectral weight is concentrated along the in-phase bands, in agreement with our qualitative reasoning. In addition, it gives the residual intensity for the out-of-phase bands, which is found to strongly depend on the hybridization with other orbitals. Indeed, we have shown that bands symmetric with respect to z will give weight along the in-phase band and vice-versa. If two bands of different symmetry with respect to z are hybridized, their weight will appear on different types of bands (in-phase or out-of-phase). For example, the $d_{xy}$ band is often significantly hybridized with $d_{xz}$/$d_{yz}$, and the out-of-phase bands then have significant weight with mainly $d_{xz}$/$d_{yz}$ character. This was called "parity switching" in ref.\cite{LinWeiKu}.

\subsection{Band symmetry}
One important consequence of defining the wave functions the way we have done is to be able to determine their symmetry. Along $k_{x}$, the important symmetry operation is the parity with respect to the $xz$ mirror plane. In Fig. 2(b), one can notice that the $xz$ plane containing the Fe is not a mirror plane (the As are not symmetric with respect to that plane). On the other hand, the one containing the As (blue line) is a true mirror plane. Therefore, the symmetry of the wave functions does not only involve the orbital character, but also the relative phase between the 2 inequivalent Fe. This situation is really a direct consequence of the existence of 2 inequivalent Fe that is induced by the As positions, as the mirror plane would contain the Fe, if there was no As. In-phase and out-of-phase bands then have opposite symmetries by construction, as sketched in Fig. 2(b) and 2(d) for $d_{xy}$ and $d_{xy}$*. This \lq\lq{}parity switching\rq\rq{} of  $d_{xy}$* also defines the symmetry of the orbitals with which it can hybridize. As we have seen that the $d_{xy}$ weight will be suppressed by unfolding effects, these hybridized orbitals will be responsible for the residual weight.

The symmetry also changes with the position in reciprocal space, as the Bloch phase term changes. This is illustrated in Fig. 2(c) and 2(e) for $d_{xy}$ and $d_{xz}$* at X, $d_{xy}$ is for example odd/yz at $\Gamma$ but even/yz at X. Here again, if the Fe were in the mirror plane, the change in phase in k-space would not change the band symmetry, so that it is a consequence of the 2 Fe situation. In Fig.\ \ref{last}(a), we sketch the electron pockets at X, coded with all features that will be important to understand the spectral weight distribution and symmetries : the main (thick lines) and folded (thin lines) characters, in the sense of the 1Fe BZ with glide-mirror symmetry [Fig. 1(c)]; the in-phase (solid lines) and out-of-phase (dotted lines) characters, as defined in II-A. The dispersion and symmetries with respect to the main mirror planes $xz$ and $yz$ are also indicated in Fig.\ \ref{last}(b) and (c). To determine if one band is in-phase or out-of-phase, we use Fig. 1(d) where the represented $d_{xy}$ (resp. $d_{xz}$/$d_{yz}$) bands are in-phase (resp. out-of-phase), as explained before. The dispersions are illustrated by the band structure of BaFe$_2$As$_2$, but the symmetry features remain the same for all structures, only the $k_z$ dependences are different. One can see that the 2 bands forming the folded electron pockets, in the sense of the 1Fe BZ with glide-mirror symmetry, $d_{yz}$  and  $d_{xy}$*, have opposite symmetries along $\Gamma$X and the same symmetries along XM. Consequently, they cross without hybridizing along $\Gamma$X, but repel each other along XM. This can only be understood after considering the symmetries derived here. As ARPES selection rules depend on the parity with respect to the mirror planes, it is also crucial to understand this to work this out to apply them correctly.

	\section{ARPES on electron pockets}

In Fig.\ \ref{FSfig}, we now use the true structure of BaFe$_2$As$_2$, whose 3D BZ is sketched in Fig.\ \ref{FSfig}(c). The only difference is that the M point in Fig. 1 is now equivalent to Z and $q$ connects different $k_z$. In Fig.\ \ref{FSfig}(a), the FS is sketched for $k_z$=1, where the FS of Fig.\ \ref{FSfig}(b) was measured (see below). Because of this, the two ellipses are now oriented in the same direction, instead of being perpendicular as in Fig. \ \ref{last} \cite{Graser2010}. Only the $k_z$ dispersion make their shape slightly different (they are slightly more rounded at $k_z$=0\cite{Graser2010}, more information on $k_z$ dependence will be given in section III-C).

\subsection{Assignment of the different bands}
\begin{figure}[tbp]
\centering
\includegraphics[width=0.45\textwidth]{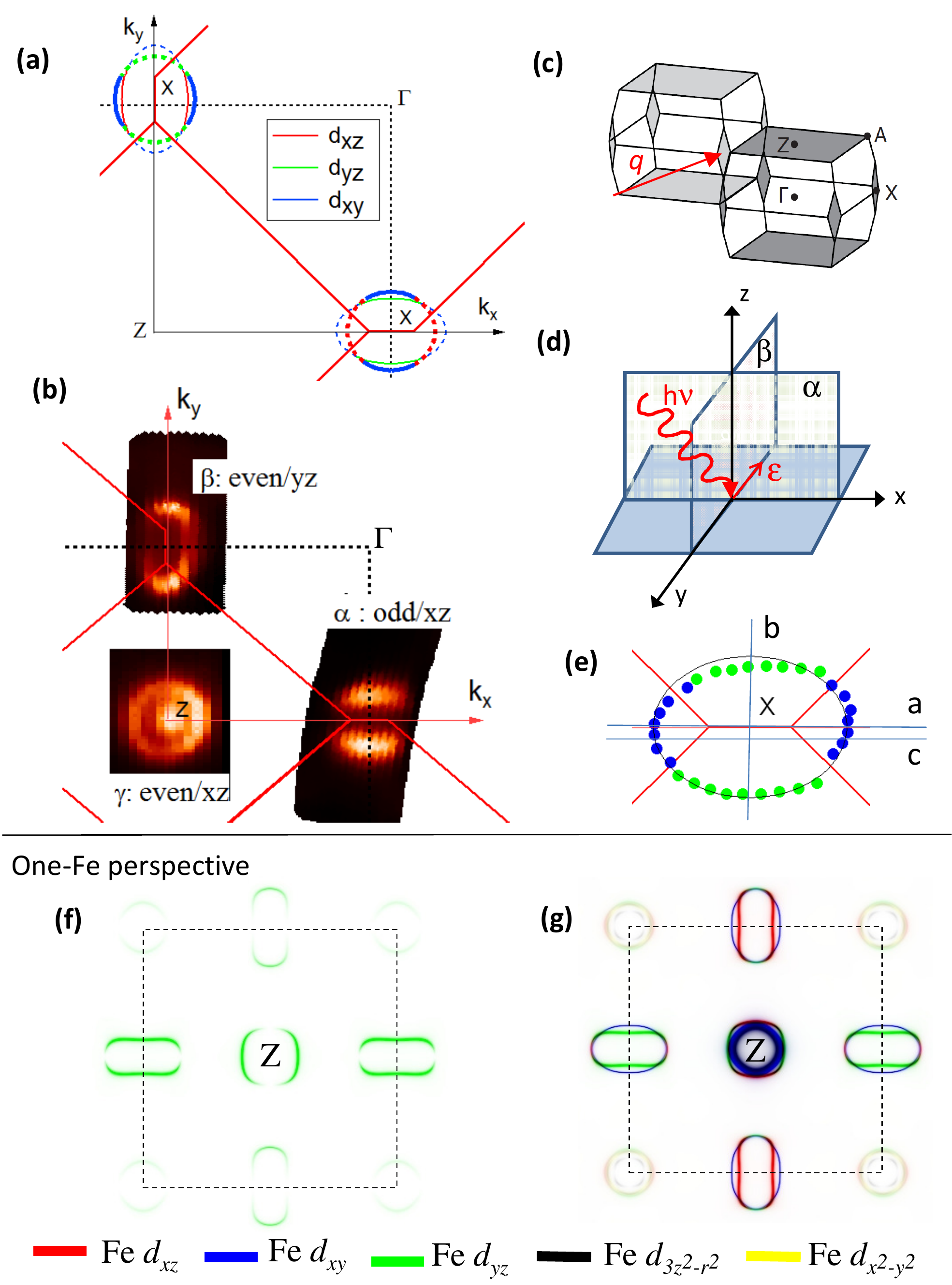}
\caption{ (a) Sketch of the Fermi Surface expected at $k_z$=1. Thick and thin lines indicate main and folded bands in the sense of the 1Fe BZ with glide-mirror symmetry. Solid and dotted lines represent the in-phase and out-of-phase orbitals. Only in-phase orbitals are expected to have sizable intensity in ARPES (see section II). (b) FS measured at T=25 K, $\hbar\omega$=34eV and polarization along $k_y$, except for the hole bands at Z, where it is in the $xz$ plane. (c) 3D BZ for BaFe$_2$As$_2$, indicating the $q$ wave vector for folding \cite{Graser2010}. (d) Geometry of the experiment. The plane of light incidence is $xz$ and the analyser slits are fixed along $k_y$. (e) FS crossings around extracted from $\alpha$ (green) and $\beta$ (blue). (f) Unfolded Fermi Surfaces at $k_z$=1 in 1Fe BZ with $d_{yz}$ spectral weight. (g) Same as (f) with all Fe 3d orbitals.}
\label{FSfig}
\end{figure}

Fig.\ \ref{FSfig}(b) presents the FS at 25K, at a photon energy $\hbar\omega$=34eV. This corresponds at normal emission to $k_z$=1  modulo $\pi$/c', where c'=6.5\AA~is the distance between 2 FeAs planes. Indeed, the photon energy $\hbar\omega$ selects one particular $k_z$, through $k_z=\sqrt{2m/\hbar^2*(\hbar\omega-W+V_0)-k_ {//}^2}$, where m is the electron mass, $k_ {//}$ the momentum in $xy$ plane, W the work function and $V_0$ an inner potential, fixed here at 14eV, as already commonly used for pnictides \cite{MalaebJPSJ09,BrouetPRB09}.

Two circular hole pockets are detected at Z, as usually observed in the literature. In most ARPES studies published so far \cite{ThirupathaiahPRB10,MalaebJPSJ09,LiuNatPhys10}, the electron pockets look like in panel $\alpha$, with the sides of the expected ellipse oriented towards Z missing. In panel $\beta$, however, we resolve clearly the parts towards Z, while the middle parts are missing. In Fig.\ \ref{FSfig}(e), we combine the FS crossings detected in these two panels (green points for $\alpha$ and blue points for $\beta$. The latter are rotated by 90$^\circ$, as $k_{x}$ and $k_y$ are equivalent by symmetry). They define a nearly perfect ellipse, matching the one sketched by thin line in Fig.\ \ref{FSfig}(a).

The difference between $\alpha$ and $\beta$ is due to the direction of light polarization, which defines the symmetry of the orbitals that can be observed. As already widely used in ARPES of iron pnictides \cite{MansartPRB11,ZhangPRB11,ThirupathaiahPRB10}, to select even or odd orbitals with respect to a mirror plane of the structure, the polarization must be even or odd with respect to that plane \cite{DamascelliRMP2003}. In our experiment, the polarization is fixed along $k_y$ and is therefore odd/$xz$ and even/$yz$ [see Fig. \ \ref{FSfig}(d)]. To view odd orbitals, we detect electrons in the $xz$ plane and, for even orbitals, we detect electrons in $yz$. This type of geometry was for example used in ref. \cite {Asensio:2003}, although it is more common to detect electrons in one plane and switch the polarization from in-plane (so-called p configuration, selecting even orbitals) to out-of-plane (s configuration for odd orbitals). For our case, the intensity of the even band is much lower in p-polarization. If Fe would belong to the mirror plane, these selection rules would only depend on the orbital character. Such a situation was implicitly assumed in all previous studies. As discussed previously, this is in fact not the case here and one should use the correct symmetries determined in Fig.\ \ref{last}(c), which change both as a function of location in the reciprocal space and of the in-phase/out-of-phase character. It turns out that all hole pockets at $\Gamma$ (or Z along 001) have in-phase characters, so that the selection rules just follow the orbital symmetry for hole pockets. However, they are completely different for electron pockets. According to the band symmetries defined in  Fig.\ \ref{last}(c), the allowed orbitals in $\alpha$ are $d_{xz}$*, $d_{xy}$ and $d_{yz}$. For $\beta$, these are $d_{xz}$*, $d_{yz}$ and $d_{xy}$*. If we rotate this panel again by 90$^\circ$ to compare it directly to $\alpha$, we just have to switch the roles of $x$ and $y$ and we expect to see orbitals even/$xz$, which are $d_{yz}$*, $d_{xz}$ and $d_{xy}$*.

%===============================
%===============================
%===============================
\begin{figure}[tbp]
\centering
\includegraphics[width=0.5\textwidth]{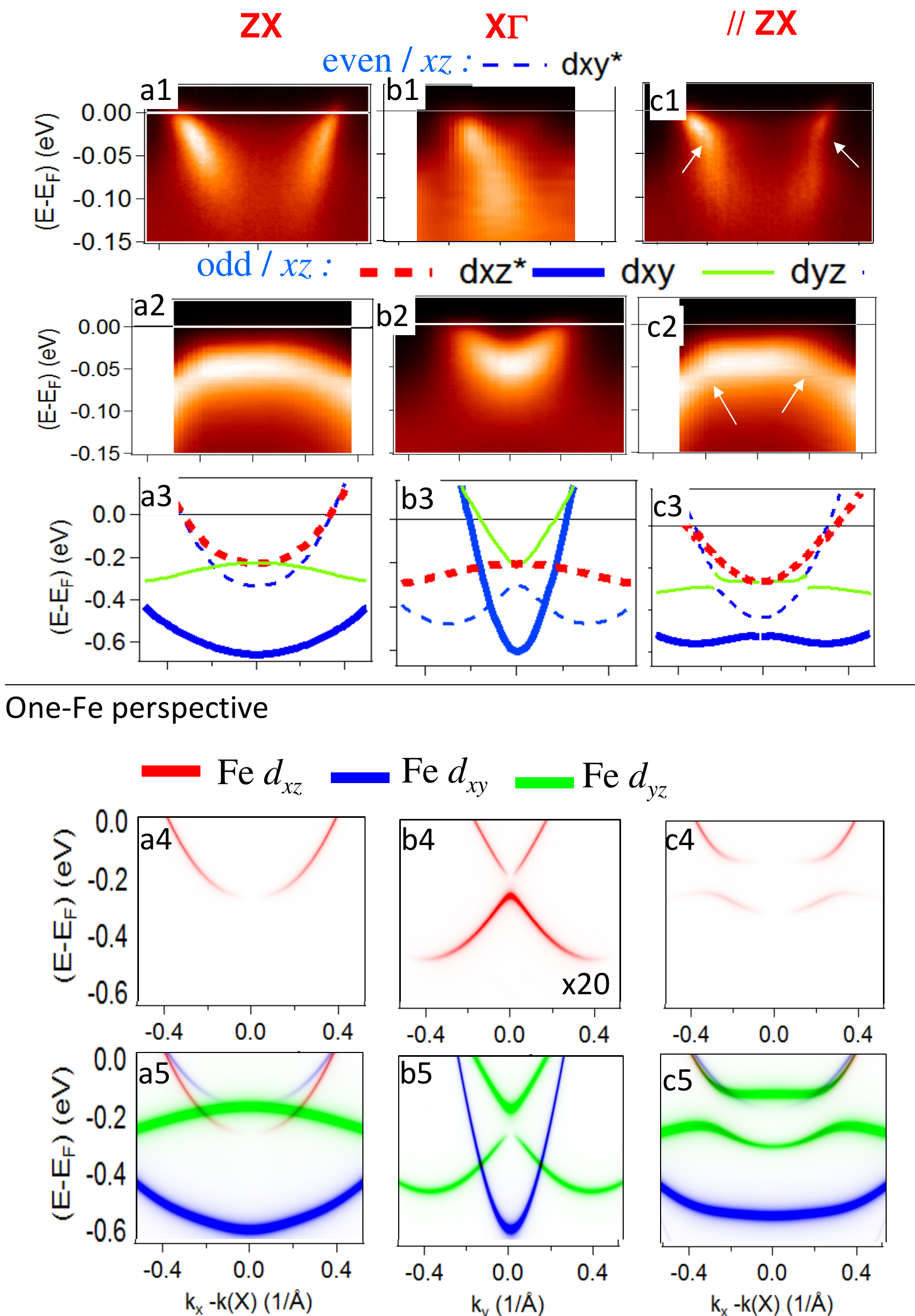}
\caption{(a-c) Dispersion observed and calculated along the 3 cuts a, b, c, indicated in Fig. \ \ref{FSfig}(e). The first row corresponds to panel $\beta$ rotated by 90$^\circ$, i.e. allowed bands are even/$xz$; the second row to $\alpha$, i.e. allowed bands are odd/$xz$; the third row to DFT calculations; the fourth and fifth rows to unfolded spectral function in 1Fe BZ with only $d_{xz}$ and all $d$ orbitals respectively. For (b4), because there is no $d_{xz}$ intensity in the original cut b, the spectral weight is plotted along a parallel $k$ path $0.04 \times \frac{2\pi}{a}$ closer to Z and enhanced by a factor of 20. For the DFT calculation (third row), the style of the lines are the same as in Fig. 3(a). We show colors of the main orbital character for clarity, although there are non-negligible hybridization effects. White arrows in (c1) and (c2) indicate kinks at the band crossings. }
\label{Disp}
\end{figure}
%===============================

In Fig.\ \ref{Disp}, we show the dispersions of the bands along the cuts indicated in Fig.\ \ref{FSfig}(e), the major ellipse axis in (a), the minor ellipse axis in (b) and an axis parallel to the major axis at $k_y$=0.05\AA$^{-1}$ in (c). For even bands, we use again panel $\beta$ rotated by 90$^\circ$. We observe that the even electron band along the major axis [Fig.\ \ref{Disp}(a1)] is very different from that along the minor axis [Fig.\ \ref{Disp}(b2)]. It is much deeper (at least 100meV) and with steeper dispersion, while the odd band is quite shallow (50meV). The calculation shows that this large  anisotropy is intrinsic to the orbital character. The shallow band is mainly formed by $d_{xz}$ or $d_{yz}$ orbitals and the deep band by the $d_{xy}$ orbital. This difference in dispersion makes it very easy to pick up the band character; it is $d_{yz}$ for the shallow band and $d_{xy}$* for the deep band. $d_{yz}$ is odd/$xz$, so that we expect to see it in $\alpha$, as we do. $d_{xy}$* is even/$xz$ and appears in $\beta$, as we expect. Note that this symmetry is opposite to that of the $d_{xy}$ orbital, so that this complete analysis is necessary to understand its parity. Because of this problem, it was wrongly assigned to $d_{x2-y2}$ in ref. \cite{ZhangPRB11}. In ref. \cite{JensenPRB11}, we correctly assigned it to $d_{xy}$, because of its dispersion, but we did not understand its parity.

Qualitatively, we predicted in section II the spectral weight to be concentrated along in-phase bands (solid lines). This explains very well the strong weight of $d_{yz}$ in (a2-c2) and also the strong suppression of $d_{xz}$*. In fact, this suppression is direct evidence for the crucial importance of these unfolding effects. We would also expect to observe clearly $d_{xy}$, for example in (b2), which is however unclear. We attribute the absence of $d_{xy}$ to its much smaller cross section for ARPES. Indeed, the smallness of polar angle of the emitted photoelectron suppresses the dipole matrix element magnitude of the purely in-plane orbitals $(d_{xy}$ and $d_{x^2-y^2})$ \cite{Hong2012}. In fact, we do observe this band weakly at other photon energies, like in Fig.\ \ref{KzFig}(c), evidencing that it is not forbidden, but only weak and masked by the intense shallow band.

On the other hand, the $d_{xy}$* band, very clear in (a1), is expected to be weak. As discussed in section II, its residual intensity is expected to be proportional to its hybridization with orbital antisymmetric with respect to $z$. In Fig.\ \ref{KzWeight}(a), we report the weight of $d_{xz}$ and $d_{xy}$ for this band, as obtained from our band calculation using Wien2k. We see that the $d_{xz}$ weight is in fact quite significant near $k_z$=1, which may explain why it is clearly seen. Interestingly, it vanishes for lower $k_z$ value, giving a way to check that the residual intensity indeed depends on this $d_{xz}$ weight. In Fig. \ \ref{KzWeight}(b), we follow the weight of this band at the Fermi level as a function of $k_z$, by changing the photon energy. The band diameter decreases away from $k_z$, as will be analyzed in III-C, but it also sharply loses intensity and disappears in $k_z$=0. This confirms very nicely the origin of this weight. The fact that it is a shadow band explains that it is not seen in most ARPES data.

Finally, it is then a combination of spectral weight distribution and matrix element effects that allows to isolate one electron ellipse. The very simple spectra allows to observe further the structure of the ellipse and how the bands couple together to form it. To understand how one goes from the deep electron band to the shallow one, it is instructive to consider the direction noted $c$, just next to the major axis. The hybridization is forbidden along $k_{x}$ because the two bands have opposite symmetries, but becomes allowed along $k_y$. Consequently, they cross without hybridizing in (a3), but, as soon as one gets away from this axis, as in (c3), a hybridization gap opens at their crossings. This forms an upper band, which progressively evolves into the shallow band of (b3). This is closely supported by our data where "kinks" signaling these hybridizations are clearly visible in (c1) and (c2) (white arrows), at the position of the two bands crossing, although we only image one band at a time.

\subsection{One-Fe perspective: unfolded band structure}

All our analysis so far are based on the 2Fe perspective, in which the phase interference between two Fe atoms in the unit cell are shown crucial to the observed ARPES spectral weight. Note however that in addition to the 2Fe basis, the same information of quasi-particle excitation can be represented in two other basis sets: the 1Fe basis with glide-mirror symmetry, and the 1Fe basis with plain translation. These three basis sets correspond to Fig.\ \ref{sketches}(c), (d) and (e) respectively, which make obvious the equivalence of the amount of information they host: compared to the 2Fe picture, the 1Fe pictures have only ``half'' of the bands, but reside in a twice larger k-space. The use of glide-mirror symmetry gives clean bands, while the use of plain translation gives more ``partial'' bands.  Figure\ \ref{sketches} should also make it obvious that all three representations respect the same sum rule and have the same total spectral weight. It is thus interesting to also consider the ARPES spectrum in an 1Fe perspective.

% Veronique : I would skip the following that was basically moved to section II

%Since the final state of the photo electron of the ARPES process correspond to a free-electron like state far away from the sample, the ARPES spectral weight is naturally given by the 1Fe pictire with a plain translation. The recently proposed band structure ``unfolding'' method~\cite{WeiKuPRL2010} provides a simple mean to calculate such an unfolded band structure from first-principles calculation. By representing the one-particle spectral function in the 1Fe Bloch basis, the above phase interference effect (associated with the Fe positions) that was hidden in the 2Fe Bloch functions now enters explicitly the theoretical spectral weight~\cite{LinWeiKu}.

Specifically, the 1Fe picture with plain translation has similar translational symmetry to the outgoing photo-electron, and would thus give a direct comparison to the ARPES spectrum. We thus proceed by unfolding the electronic structure in the 1Fe basis with plain translation.  Since the basis consists of a single Fe atom in the unit cell, the unfolded band structure absorbs the structure related phase interference, leaving only a trivial orbital dependent dipole matrix element of a \textit{single} Fe atom, tunable by the polarization of the incident light.

%Since the basis now consists of a single Fe atom in the unit cell, a direct comparison with the experimental spectral weight can therefore be easily made with the help of simply an additional dipole matrix elements of a \textit{single} Fe. That is, the unfolded band structure automatically includes all the phase interference effects discussed above, and gives directly the ARPES spectral weight after being multiplied by an orbital dependent dipole matrix element of a \textit{single} Fe atom.

% Veronique : I don\rq{}t understand the following sentence. Rotation is as valid in 1FeBZ than 2Fe BZ...
%Note that since the unfolded band structure give direct access to the larger 1Fe BZ, in the following discussion we will compare directly to the observed raw spectrum around ($\pi$,0,$\pi$) and  (0,$\pi$,$\pi$) without rotating the information into the same pocket at ($\pi$,0,$\pi$) as was done in the above assignment.

%===============================
\begin{figure}[tbp]
\centering
\includegraphics[width=0.48\textwidth]{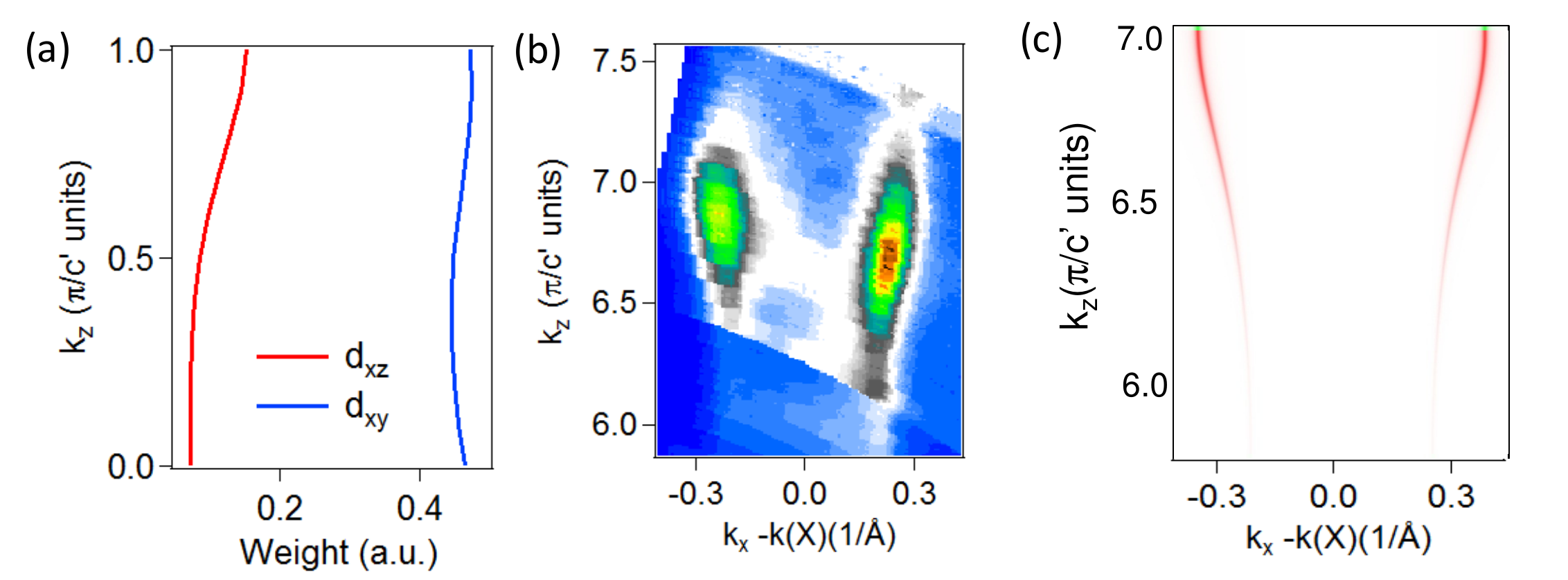}
%{Fig5_KzWeight.jpg}
\caption{(a) Weight for $d_{xy}$ and $d_{xz}$ in the deep electron band as a function of $k_z$ obtained in the calculation. (b) Intensity of the even deep electron band along ZX integrated at the Fermi level, as a function of $k_{z}$. (c) Unfolded spectral function for the deep electron band shown in (b).}
\label{KzWeight}
\end{figure}
%===============================

%===============================
%\begin{figure}[tbp]
%\centering
%%\includegraphics[width=0.48\textwidth]{Fig6_FSsimul.jpg}
%\includegraphics[width=0.48\textwidth]{fig6.pdf}
%\caption{(a) Unfolded Fermi Surfaces in 1Fe BZ with spectral weights including all Fe $d$ orbitals. (b) Only $d_{yz}$ character is shown.}
%\label{FSsimul}
%\end{figure}
%===============================

Figure\ \ref{FSfig}(g) shows the calculated unfolded Fermi surfaces in the 1Fe BZ, colored according to the orbital character and with intensity proportional to the calculated spectral weight. This result is obviously very different from the 1Fe picture under the glide-mirror symmetry (c.f. Fig.\ \ref{sketches}(d)), but is in good agreement with the expected spectral distribution in Fig.~\ref{sketches}(e). Given that our experimental polarization is along the $y$-direction, the atomic matrix element would thus be dominated by the $yz$ orbital. Indeed, Fig.\ \ref{FSfig}(f) gives the Fermi surface weighted by only the $yz$ character and appears to reproduce very well the experimental observation in Fig.\ \ref{FSfig}(b).

The power and simplicity of the unfolding picture is further illustrated in the last two rows of Fig.\ \ref{Disp}, which follow the similar $k$-paths as those of the first two rows. Let's first focus on the last row (a5-c5) and only pay attention to the $yz$ (green) character that contains very strong spectral weight. Evidently, all our experimentally observed features in the 2nd row (a2-c2) are in excellent agreement with the unfolded band structure. Interestingly, the calculation also establishes that the ARPES feature observed in (c2) is actually not from a single band, but from two crossing bands, as hinted by the observed ``break'' in the dispersion [indicated by the arrows in (c2)]. A similarly excellent agreement is found between the fourth row (a4-c4) and the first row (a1-c1). Particularly, the same ``break'' in the dispersion [indicated by arrows in (c1)] can be resolved from the two crossing bands. [For clarity and to be consistent with the analysis in the previous section, (a4-c4) shows only the weak intensity of the $xz$ character on the electron pocket around ($\pi$,0,$\pi$), which gives the same intensity as the experimentally observed $yz$ character around ($\pi$,0,$\pi$).]

%WK: I don't think it is necessary to explain the feature at higher binding energy, as even the theory has a vanishing weight near the X point.

%Along X$\Gamma$, the simulation predicts a strong $d_{yz}$ band, as observed in (b2), a strong suppression of $d_{xz}$*, consistent with (b2), and also a clear $d_{xy}$, which is not clearly detected. It is odd/$xz$ and should appear in (b2). We attribute the absence of $d_{xy}$ to its much smaller cross section for ARPES. Indeed, the smallness of polar angle of the emitted photoelectron suppresses the dipole matrix element magnitude of the purely in-plane orbitals $(d_{xy}$ and $d_{x^2-y^2})$ \cite{Hong2012}. In fact, we do observe this band weakly at other photon energies, like in Fig.\ \ref{KzFig}(c), evidencing that it is not forbidden, but only weak and masked by the intense shallow band.

Notice that exactly along the path corresponding to Fig.\ \ref{Disp}(b1), there is practically no theoretical spectral weight from any of the even unfolded bands. %As shown in a recent study~\cite{LinWeiKu}, this lack of spectral weight turns out to reflect the important fact that the electron pockets are actually created by the breaking of the 1Fe translational symmetry.
In our experimental spectrum, however, a very weak band can be recognized. A careful analysis of the unfolded band structure indicate that this very weak feature most likely originates from a finite momentum resolution of ARPES. Indeed, Fig.\ \ref{Disp}(b4) shows that slightly away from the path, a weak spectral weight appears that resemble the experimental observation. (An energy-broadening of the theoretical spectrum is necessary to make a proper comparison.)

%Notice that the very weak feature in (b1) has no $yz$ character and actually originates from the $xz$ character. [Figure\ \ref{Disp}(b4) gives $xz$ character as well.] This is an interesting observation that reflects an important (but mostly unnoticed) characteristic of the low-energy Fe-d Wannier orbitals. As shown in the previous first-principles calculation~\cite{Lee2009}, the low-energy Wannier orbitals in the tetrogonal coordination actually contains $p$-character as well, due to their hybridization with the As orbitals. Specifically, the $xz$ orbital contains some $p_y$ character as evident from the tail of the Wannier orbital bending toward the $y$ direction, but with opposite sign across the $xz$ plane. It is in fact this hybridization that changes completely/qualitatively the hopping integrals and the corresponding ferro-orbital and anti-ferromagnetic correlation~\cite{Lee2009}. This $p_y$ character gives the $xz$ orbital a finite dipole matrix element and thus makes visible the band in (b1).

%The comparison with (a3-c3) confirms the previous analysis, as all in-phase bands have strong weight and the out-of-phase bands no or weaker weight, with a systematic change of color, indicating parity switching.

It is also instructive to compare the theory in the 2Fe picture in Fig.\ \ref{Disp}(a3-c3) and the 1Fe picture in (a5-c5). They confirm the previous analysis, as all in-phase bands have strong weight and the out-of-phase bands no or weaker weight with a different color. Of particular interest is the character of the electron pockets in (a3) and (a5). As shown in a recent study~\cite{LinWeiKu}, the reduced spectral weight at the Fermi level in this direction reflects the important fact that the electron pockets are actually created by the breaking of the 1Fe translational symmetry. The weak (red and blue) electron pockets in Fig.\ \ref{Disp}(a5) \textit{appear} to have the opposite character as the regular assignment in the 2Fe picture in (a3). For example, the lighter $xy$ (blue) band in the 2Fe picture \textit{appears} to be heavier than the $xz$ (red) band in the 1Fe picture. Since the ARPES spectral weight will coincide with the unfolded band structure in (a5), an experimental analysis of the polarization dependence might mistakenly assign the heavier pocket as the $xy$ pocket in the 2Fe picture, seemingly in contradiction with the theoretical results in (a3). Based on such comparison, one might even draw the wrong conclusion of a strongly orbital dependent mass renormalization. However, this switch of character is exactly to be expected from the unfolding theory. As recently shown~\cite{LinWeiKu}, the folding potential here originates from the alternating out-of-plane positioning of As atoms, and thus only couples orbitals with opposite z-parity. Thus, the folding of band structures is always ``parity switching''. Therefore, a $xz$ pocket near (0,$\pi$,0) would be folded to ($\pi$,0,$\pi$) and switch to $xy$ character instead. (The same ``parity switch'' would apply to the other $xy$ pocket as well to give the folded pocket $xz$ character.) This parity switching characteristic \textit{has to} be taken into account when assigning band character in ARPES analysis, and highlights the convenience of the theoretical unfolded band structure in assisting ARPES analysis.

Finally, not surprisingly, the $k_z$ dependence of the spectral weight can also be captured. Figure\ \ref{KzWeight}(c) shows that the growing $xz$ character near $k_z=\pi$, and consequently a strongly enhanced ARPES spectral weight, in perfect agreement with the experiment in Figure\ \ref{KzWeight}(b). It should be obvious that the unfolded band structure and Fermi surfaces offer a simple and direct way to analyze ARPES spectral function in general. It is particularly valuable for the case of families of Fe-based superconductors, due to the strongly k-dependent orbital character of the bands, and the special ``parity switching'' nature of the band folding~\cite{LinWeiKu}.

%===============================
\begin{figure}[tbp]
\centering
\includegraphics[width=0.48\textwidth]{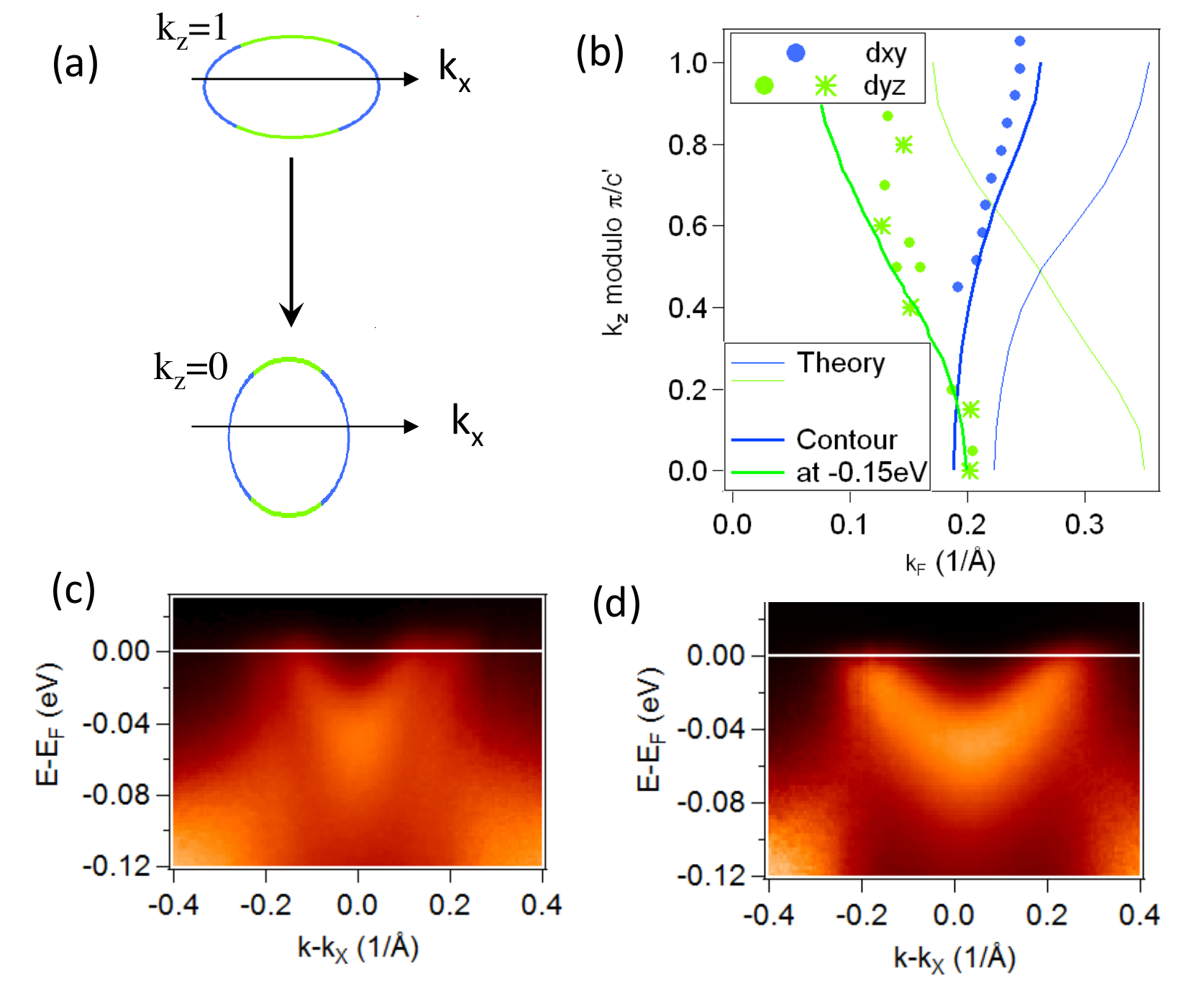}
\caption{(a) Evolution of the electron pocket evolution expected as a function of kz. (b) Size of the electron pockets as a function of $k_z$. Points indicate $k_F$ along $k_{x}$ for the deep electron band (blue points) and along $k_y$ for the shallow one (green points). The points for the deep band were extracted from Fig. 5(c). For the shallow electron band, two samples were measured for 20<$\hbar\omega$<45eV (circles) and 40<$\hbar\omega$<90eV (stars). Thick lines correspond to values expected in calculations for BaFe$_2$As$_2$ doped by 10$\%$ Co. Thin lines are the contours calculated at -0.15eV. (c) Shallow electron band at 39eV ($k_{z}$[X]=7) (d) Shallow electron band at 27eV ($k_{z}$[X]=6).}
\label{KzFig}
\end{figure}
%===============================

\subsection{$k_z$ dispersion}

To completely characterize the band structure, we should follow the dispersion as a function of $k_z$. Theoretically, the ellipse major axis is expected to rotate by 90$^{\circ}$ between $k_z=1$ and  $k_z=0$. This is required to have a 4$_2$/m axis at X. This is obtained by squeezing the pockets, as sketched in Fig.\ \ref{KzFig}(a) for one pocket \cite{Graser2010}. Note that the orbital character does not rotate, so that the asymmetry in $v_F$ between $d_{xy}$ and $d_{xz}$/$d_{yz}$ remains the same at both $k_z$. As the folded pocket has opposite characters ($d_{xz}$ along $k_x$ and $d_{xy}$ along $k_y$), the superposition of the 2 pockets is needed to obtain the equivalency between the 2 perpendicular directions.

Fig.\ \ref{KzWeight}(b) shows that the deep electron band diameter indeed decreases away from $k_z=1$ and Fig.\ \ref{KzFig}(c) and (d) show that the shallow band diameter significantly expands between photon energies corresponding to $k_z$=1 and 0. In Fig.\ \ref{KzFig}(b), we report by points $k_F$ as a function of $k_z$ for the deep electron band (measured along $k_x$) and the shallow one (measured along $k_y$). We compare these points to those expected in the calculation (thin lines). Qualitatively, the expected evolution is there, but it is quantitatively much weaker. In fact, both bands form significantly smaller pockets at all $k_{z}$ than the calculated ones; $k_F$ is nearly a factor 2 smaller. This tendency was already noted both by de Haas-Van Alphen measurements in other compounds \cite{ColdeaPRL08} and ARPES \cite{BrouetPRB09}. We observe that the contours calculated at -0.15eV for both bands (thick lines) are in much better agreement with our data, supporting the idea that such a very large shift is necessary to adjust the data. It is very important to define this shift accurately to estimate the renormalization of the bands. With this shift, the warping expected along $k_z$ corresponds quite well to the calculated one.

\section{Conclusion}

Finally, we obtained very simple spectra for the electron pockets, despite the complexity of the band structure. Only one band is detected in each measurement, $d_{yz}$ in the odd configuration and $d_{xy}$* in the even one. These two bands combine to form one full ellipse in the sense of the 1Fe BZ of Fig. 1(d), as sketched in Fig. 3(e). To our knowledge, this is the first time that a complete ellipse can be clearly resolved by ARPES. We explain why the other ellipse, expected to be formed by $d_{xz}$*/$d_{xy}$, is largely suppressed in our measurements by a combination of spectral weight distribution (suppressing strongly $d_{xz}$*) and matrix element effects (suppressing very efficiently $d_{xy}$). To clarify this point, we consider explicitly the spectral weight distribution expected from interferences between the 2 Fe of the unit cell and the associated parity switching effects. These effects were neglected in previous ARPES studies, but they are found very important to correctly interpret spectra near X points. Our data are in excellent agreement both with our qualitative symmetry arguments and with direct comparison with the unfolded band structure of ref. \cite{LinWeiKu}. We note that some authors prefer to describe electron pockets as one outer sheet (mainly of $d_{xy}$ character) and one inner sheet (mainly of  $d_{xz}$/ $d_{yz}$ character) rather than as 2 crossed ellipses. The two descriptions can be consistent with our analysis, provided the gap between the two sheets remains quite small, as our measurements rule out a strong deviation of the electron pockets from the elliptic shape. We emphasize that, in any case, the spectral weight will be strongly modulated around these sheets, which will affect all electronic properties and disrupts the notion of sheet.

Resolving one full ellipse is a favorable experimental situation, as we can then study in details the properties of each bands, how they couple together and how they evolve with $k_z$. As the other ellipse can be obtained by folding, there is no information missing about the electronic structure of the electron pockets. We confirm unambiguously the orbital characters predicted by the band structure calculations for the electron pockets and a large anisotropy in Fermi velocities, the $d_{xy}$ band having a much steeper dispersion than $d_{xz}$/ $d_{yz}$. We detect a $k_{z}$ dependence of the pockets qualitatively consistent with theory, although their sizes itself is much smaller. The calculated bands must be shifted by 0.15eV to match the experiment.

More generally, our study allows to get a better understanding of how to treat folding effects in ARPES. This should be useful for further studies of pnictides and in all, rather common, situations where the unit cell contains 2 inequivalent atoms or more. We show that the spectral weight distribution does not follow the 2Fe BZ scheme, nor that commonly referred to as the 1Fe BZ scheme \cite{MazinPRL08,Graser2010,Andersen2011}. The key to understand the folding here is to consider separately bands having the symmetry of the folding potential or the opposite symmetry. An important consequence we derive and confirm experimentally is that the symmetry of the bands are not on-site properties anymore, but that they depend on the location in k-space and on the nature of the band (main or folded).

Furthermore, the distribution of spectral weight is an intrinsic feature of the electronic structure that defines its symmetry and can have important consequences. For example, the FS possesses a 4-fold rotation axis at the X point in the 2Fe BZ description (Fig. 1c, this becomes a 4$_2$/m screw axis in the 3D structure of BaFe$_2$As$_2$), but only a 2-fold rotation axis in the 1Fe BZ (Fig. 1d). Which is the correct symmetry in iron pnictides ? Recently, the anisotropy of spin fluctuations measured by neutrons at the X point were attributed to the true symmetry of the underlying Fe lattice that misses the 4-fold axis \cite{ParkPRB10}. To what extent this "hidden" anisotropy can play a role in the various "anisotropies" and "nemacities" observed in these materials is still an open question. Our ARPES study clearly evidence the spectral weight modulation corresponding to the folding, implying there is no true 4$_2$/m axis at the X point.

\section{Acknowledgements}
We thank I.I. Mazin and D. Malterre for decisive comments. We acknowledge useful discussions with M. Aichhorn, S. Biermann, M.H. Fisher, S. Graser, M. Grioni, P. Hirshfeld, A.F. Kemper and F. Rullier-Albenque. Financial support from the French RTRA Triangle de la physique and the ANR \lq\lq{}Pnictides\rq\rq{} is acknowledged.

\bibliography{Bib}

%merlin.mbs apsrev4-1.bst 2010-07-25 4.21a (PWD, AO, DPC) hacked
%Control: key (0)
%Control: author (72) initials jnrlst
%Control: editor formatted (1) identically to author
%Control: production of article title (-1) disabled
%Control: page (0) single
%Control: year (1) truncated
%Control: production of eprint (0) enabled
\begin{thebibliography}{28}%
\makeatletter
\providecommand \@ifxundefined [1]{%
 \@ifx{#1\undefined}
}%
\providecommand \@ifnum [1]{%
 \ifnum #1\expandafter \@firstoftwo
 \else \expandafter \@secondoftwo
 \fi
}%
\providecommand \@ifx [1]{%
 \ifx #1\expandafter \@firstoftwo
 \else \expandafter \@secondoftwo
 \fi
}%
\providecommand \natexlab [1]{#1}%
\providecommand \enquote  [1]{``#1''}%
\providecommand \bibnamefont  [1]{#1}%
\providecommand \bibfnamefont [1]{#1}%
\providecommand \citenamefont [1]{#1}%
\providecommand \href@noop [0]{\@secondoftwo}%
\providecommand \href [0]{\begingroup \@sanitize@url \@href}%
\providecommand \@href[1]{\@@startlink{#1}\@@href}%
\providecommand \@@href[1]{\endgroup#1\@@endlink}%
\providecommand \@sanitize@url [0]{\catcode `\\12\catcode `\$12\catcode
  `\&12\catcode `\#12\catcode `\^12\catcode `\_12\catcode `\%12\relax}%
\providecommand \@@startlink[1]{}%
\providecommand \@@endlink[0]{}%
\providecommand \url  [0]{\begingroup\@sanitize@url \@url }%
\providecommand \@url [1]{\endgroup\@href {#1}{\urlprefix }}%
\providecommand \urlprefix  [0]{URL }%
\providecommand \Eprint [0]{\href }%
\providecommand \doibase [0]{http://dx.doi.org/}%
\providecommand \selectlanguage [0]{\@gobble}%
\providecommand \bibinfo  [0]{\@secondoftwo}%
\providecommand \bibfield  [0]{\@secondoftwo}%
\providecommand \translation [1]{[#1]}%
\providecommand \BibitemOpen [0]{}%
\providecommand \bibitemStop [0]{}%
\providecommand \bibitemNoStop [0]{.\EOS\space}%
\providecommand \EOS [0]{\spacefactor3000\relax}%
\providecommand \BibitemShut  [1]{\csname bibitem#1\endcsname}%
\let\auto@bib@innerbib\@empty
%</preamble>
\bibitem [{\citenamefont {Mazin}\ \emph {et~al.}(2008)\citenamefont {Mazin},
  \citenamefont {Singh}, \citenamefont {Johannes},\ and\ \citenamefont
  {Du}}]{MazinPRL08}%
  \BibitemOpen
  \bibfield  {author} {\bibinfo {author} {\bibfnamefont {I.~I.}\ \bibnamefont
  {Mazin}}, \bibinfo {author} {\bibfnamefont {D.~J.}\ \bibnamefont {Singh}},
  \bibinfo {author} {\bibfnamefont {M.~D.}\ \bibnamefont {Johannes}}, \ and\
  \bibinfo {author} {\bibfnamefont {M.~H.}\ \bibnamefont {Du}},\ }\href@noop {}
  {\bibfield  {journal} {\bibinfo  {journal} {Phys. Rev. Lett.}\ }\textbf
  {\bibinfo {volume} {101}},\ \bibinfo {pages} {057003} (\bibinfo {year}
  {2008})}\BibitemShut {NoStop}%
\bibitem [{\citenamefont {Paglione}\ and\ \citenamefont
  {Greene}(2010)}]{Paglione:2010p271}%
  \BibitemOpen
  \bibfield  {author} {\bibinfo {author} {\bibfnamefont {J.}~\bibnamefont
  {Paglione}}\ and\ \bibinfo {author} {\bibfnamefont {R.~L.}\ \bibnamefont
  {Greene}},\ }\href@noop {} {\bibfield  {journal} {\bibinfo  {journal} {Nature
  Phys.}\ }\textbf {\bibinfo {volume} {6}},\ \bibinfo {pages} {645} (\bibinfo
  {year} {2010})}\BibitemShut {NoStop}%
\bibitem [{\citenamefont {Thirupathaiah}\ \emph {et~al.}(2010)\citenamefont
  {Thirupathaiah}, \citenamefont {de~Jong}, \citenamefont {Ovsyannikov},
  \citenamefont {D\"urr}, \citenamefont {Varykhalov}, \citenamefont {Follath},
  \citenamefont {Huang}, \citenamefont {Huisman}, \citenamefont {Golden},
  \citenamefont {Zhang}, \citenamefont {Jeschke}, \citenamefont {Valent\'\i},
  \citenamefont {Erb}, \citenamefont {Gloskovskii},\ and\ \citenamefont
  {Fink}}]{ThirupathaiahPRB10}%
  \BibitemOpen
  \bibfield  {author} {\bibinfo {author} {\bibfnamefont {S.}~\bibnamefont
  {Thirupathaiah}}, \bibinfo {author} {\bibfnamefont {S.}~\bibnamefont
  {de~Jong}}, \bibinfo {author} {\bibfnamefont {R.}~\bibnamefont
  {Ovsyannikov}}, \bibinfo {author} {\bibfnamefont {H.~A.}\ \bibnamefont
  {D\"urr}}, \bibinfo {author} {\bibfnamefont {A.}~\bibnamefont {Varykhalov}},
  \bibinfo {author} {\bibfnamefont {R.}~\bibnamefont {Follath}}, \bibinfo
  {author} {\bibfnamefont {Y.}~\bibnamefont {Huang}}, \bibinfo {author}
  {\bibfnamefont {R.}~\bibnamefont {Huisman}}, \bibinfo {author} {\bibfnamefont
  {M.~S.}\ \bibnamefont {Golden}}, \bibinfo {author} {\bibfnamefont {Y.-Z.}\
  \bibnamefont {Zhang}}, \bibinfo {author} {\bibfnamefont {H.~O.}\ \bibnamefont
  {Jeschke}}, \bibinfo {author} {\bibfnamefont {R.}~\bibnamefont {Valent\'\i}},
  \bibinfo {author} {\bibfnamefont {A.}~\bibnamefont {Erb}}, \bibinfo {author}
  {\bibfnamefont {A.}~\bibnamefont {Gloskovskii}}, \ and\ \bibinfo {author}
  {\bibfnamefont {J.}~\bibnamefont {Fink}},\ }\href@noop {} {\bibfield
  {journal} {\bibinfo  {journal} {Phys. Rev. B}\ }\textbf {\bibinfo {volume}
  {81}},\ \bibinfo {pages} {104512} (\bibinfo {year} {2010})}\BibitemShut
  {NoStop}%
\bibitem [{\citenamefont {Malaeb}\ \emph {et~al.}(2009)\citenamefont {Malaeb},
  \citenamefont {Yoshida}, \citenamefont {Fujimori}, \citenamefont {Kubota},
  \citenamefont {Ono}, \citenamefont {Kihou}, \citenamefont {Shirage},
  \citenamefont {Kito}, \citenamefont {Iyo}, \citenamefont {Eisaki},
  \citenamefont {Nakajima}, \citenamefont {Tamegai},\ and\ \citenamefont
  {Arita}}]{MalaebJPSJ09}%
  \BibitemOpen
  \bibfield  {author} {\bibinfo {author} {\bibfnamefont {W.}~\bibnamefont
  {Malaeb}}, \bibinfo {author} {\bibfnamefont {T.}~\bibnamefont {Yoshida}},
  \bibinfo {author} {\bibfnamefont {A.}~\bibnamefont {Fujimori}}, \bibinfo
  {author} {\bibfnamefont {M.}~\bibnamefont {Kubota}}, \bibinfo {author}
  {\bibfnamefont {K.}~\bibnamefont {Ono}}, \bibinfo {author} {\bibfnamefont
  {K.}~\bibnamefont {Kihou}}, \bibinfo {author} {\bibfnamefont {P.~M.}\
  \bibnamefont {Shirage}}, \bibinfo {author} {\bibfnamefont {H.}~\bibnamefont
  {Kito}}, \bibinfo {author} {\bibfnamefont {A.}~\bibnamefont {Iyo}}, \bibinfo
  {author} {\bibfnamefont {H.}~\bibnamefont {Eisaki}}, \bibinfo {author}
  {\bibfnamefont {Y.}~\bibnamefont {Nakajima}}, \bibinfo {author}
  {\bibfnamefont {T.}~\bibnamefont {Tamegai}}, \ and\ \bibinfo {author}
  {\bibfnamefont {R.}~\bibnamefont {Arita}},\ }\href@noop {} {\bibfield
  {journal} {\bibinfo  {journal} {Journal of the Physical Society of Japan}\
  }\textbf {\bibinfo {volume} {78}},\ \bibinfo {pages} {123706} (\bibinfo
  {year} {2009})}\BibitemShut {NoStop}%
\bibitem [{\citenamefont {Liu}\ \emph {et~al.}(2010)\citenamefont {Liu},
  \citenamefont {Kondo}, \citenamefont {Fernandes}, \citenamefont {Palczewski},
  \citenamefont {Mun}, \citenamefont {Ni}, \citenamefont {Thaler},
  \citenamefont {Bostwick}, \citenamefont {Rotenberg}, \citenamefont
  {Schmalian}, \citenamefont {Bud’ko}, \citenamefont {Canfield},\ and\
  \citenamefont {Kaminski}}]{LiuNatPhys10}%
  \BibitemOpen
  \bibfield  {author} {\bibinfo {author} {\bibfnamefont {C.}~\bibnamefont
  {Liu}}, \bibinfo {author} {\bibfnamefont {T.}~\bibnamefont {Kondo}}, \bibinfo
  {author} {\bibfnamefont {R.}~\bibnamefont {Fernandes}}, \bibinfo {author}
  {\bibfnamefont {A.}~\bibnamefont {Palczewski}}, \bibinfo {author}
  {\bibfnamefont {E.}~\bibnamefont {Mun}}, \bibinfo {author} {\bibfnamefont
  {N.}~\bibnamefont {Ni}}, \bibinfo {author} {\bibfnamefont {A.}~\bibnamefont
  {Thaler}}, \bibinfo {author} {\bibfnamefont {A.}~\bibnamefont {Bostwick}},
  \bibinfo {author} {\bibfnamefont {E.}~\bibnamefont {Rotenberg}}, \bibinfo
  {author} {\bibfnamefont {J.}~\bibnamefont {Schmalian}}, \bibinfo {author}
  {\bibfnamefont {S.}~\bibnamefont {Bud’ko}}, \bibinfo {author}
  {\bibfnamefont {P.}~\bibnamefont {Canfield}}, \ and\ \bibinfo {author}
  {\bibfnamefont {A.}~\bibnamefont {Kaminski}},\ }\href@noop {} {\bibfield
  {journal} {\bibinfo  {journal} {Nature Physics}\ }\textbf {\bibinfo {volume}
  {6}},\ \bibinfo {pages} {419} (\bibinfo {year} {2010})}\BibitemShut {NoStop}%
\bibitem [{\citenamefont {Zhang}\ \emph {et~al.}(2011)\citenamefont {Zhang},
  \citenamefont {Chen}, \citenamefont {He}, \citenamefont {Zhou}, \citenamefont
  {Xie}, \citenamefont {Fang}, \citenamefont {Tsai}, \citenamefont {Chen},
  \citenamefont {Hayashi}, \citenamefont {Jiang}, \citenamefont {Iwasawa},
  \citenamefont {Shimada}, \citenamefont {Namatame}, \citenamefont {Taniguchi},
  \citenamefont {Hu},\ and\ \citenamefont {Feng}}]{ZhangPRB11}%
  \BibitemOpen
  \bibfield  {author} {\bibinfo {author} {\bibfnamefont {Y.}~\bibnamefont
  {Zhang}}, \bibinfo {author} {\bibfnamefont {F.}~\bibnamefont {Chen}},
  \bibinfo {author} {\bibfnamefont {C.}~\bibnamefont {He}}, \bibinfo {author}
  {\bibfnamefont {B.}~\bibnamefont {Zhou}}, \bibinfo {author} {\bibfnamefont
  {B.~P.}\ \bibnamefont {Xie}}, \bibinfo {author} {\bibfnamefont
  {C.}~\bibnamefont {Fang}}, \bibinfo {author} {\bibfnamefont {W.~F.}\
  \bibnamefont {Tsai}}, \bibinfo {author} {\bibfnamefont {X.~H.}\ \bibnamefont
  {Chen}}, \bibinfo {author} {\bibfnamefont {H.}~\bibnamefont {Hayashi}},
  \bibinfo {author} {\bibfnamefont {J.}~\bibnamefont {Jiang}}, \bibinfo
  {author} {\bibfnamefont {H.}~\bibnamefont {Iwasawa}}, \bibinfo {author}
  {\bibfnamefont {K.}~\bibnamefont {Shimada}}, \bibinfo {author} {\bibfnamefont
  {H.}~\bibnamefont {Namatame}}, \bibinfo {author} {\bibfnamefont
  {M.}~\bibnamefont {Taniguchi}}, \bibinfo {author} {\bibfnamefont {J.~P.}\
  \bibnamefont {Hu}}, \ and\ \bibinfo {author} {\bibfnamefont {D.~L.}\
  \bibnamefont {Feng}},\ }\href@noop {} {\bibfield  {journal} {\bibinfo
  {journal} {Phys. Rev. B}\ }\textbf {\bibinfo {volume} {83}},\ \bibinfo
  {pages} {054510} (\bibinfo {year} {2011})}\BibitemShut {NoStop}%
\bibitem [{\citenamefont {Rullier-Albenque}\ \emph {et~al.}(2009)\citenamefont
  {Rullier-Albenque}, \citenamefont {Colson}, \citenamefont {Forget},\ and\
  \citenamefont {Alloul}}]{RullierAlbenquePRL09}%
  \BibitemOpen
  \bibfield  {author} {\bibinfo {author} {\bibfnamefont {F.}~\bibnamefont
  {Rullier-Albenque}}, \bibinfo {author} {\bibfnamefont {D.}~\bibnamefont
  {Colson}}, \bibinfo {author} {\bibfnamefont {A.}~\bibnamefont {Forget}}, \
  and\ \bibinfo {author} {\bibfnamefont {H.}~\bibnamefont {Alloul}},\
  }\href@noop {} {\bibfield  {journal} {\bibinfo  {journal} {Phys. Rev. Lett.}\
  }\textbf {\bibinfo {volume} {103}},\ \bibinfo {pages} {057001} (\bibinfo
  {year} {2009})}\BibitemShut {NoStop}%
\bibitem [{\citenamefont {Fang}\ \emph {et~al.}(2009)\citenamefont {Fang},
  \citenamefont {Luo}, \citenamefont {Cheng}, \citenamefont {Wang},
  \citenamefont {Jia}, \citenamefont {Mu}, \citenamefont {Shen}, \citenamefont
  {Mazin}, \citenamefont {Shan}, \citenamefont {Ren},\ and\ \citenamefont
  {Wen}}]{FangPRB09}%
  \BibitemOpen
  \bibfield  {author} {\bibinfo {author} {\bibfnamefont {L.}~\bibnamefont
  {Fang}}, \bibinfo {author} {\bibfnamefont {H.}~\bibnamefont {Luo}}, \bibinfo
  {author} {\bibfnamefont {P.}~\bibnamefont {Cheng}}, \bibinfo {author}
  {\bibfnamefont {Z.}~\bibnamefont {Wang}}, \bibinfo {author} {\bibfnamefont
  {Y.}~\bibnamefont {Jia}}, \bibinfo {author} {\bibfnamefont {G.}~\bibnamefont
  {Mu}}, \bibinfo {author} {\bibfnamefont {B.}~\bibnamefont {Shen}}, \bibinfo
  {author} {\bibfnamefont {I.~I.}\ \bibnamefont {Mazin}}, \bibinfo {author}
  {\bibfnamefont {L.}~\bibnamefont {Shan}}, \bibinfo {author} {\bibfnamefont
  {C.}~\bibnamefont {Ren}}, \ and\ \bibinfo {author} {\bibfnamefont {H.-H.}\
  \bibnamefont {Wen}},\ }\href {\doibase 10.1103/PhysRevB.80.140508} {\bibfield
   {journal} {\bibinfo  {journal} {Phys. Rev. B}\ }\textbf {\bibinfo {volume}
  {80}},\ \bibinfo {pages} {140508} (\bibinfo {year} {2009})}\BibitemShut
  {NoStop}%
\bibitem [{\citenamefont {Chauvi\`ere}\ \emph {et~al.}(2010)\citenamefont
  {Chauvi\`ere}, \citenamefont {Gallais}, \citenamefont {Cazayous},
  \citenamefont {M\'easson}, \citenamefont {Sacuto}, \citenamefont {Colson},\
  and\ \citenamefont {Forget}}]{ChauvierePRB10}%
  \BibitemOpen
  \bibfield  {author} {\bibinfo {author} {\bibfnamefont {L.}~\bibnamefont
  {Chauvi\`ere}}, \bibinfo {author} {\bibfnamefont {Y.}~\bibnamefont
  {Gallais}}, \bibinfo {author} {\bibfnamefont {M.}~\bibnamefont {Cazayous}},
  \bibinfo {author} {\bibfnamefont {M.~A.}\ \bibnamefont {M\'easson}}, \bibinfo
  {author} {\bibfnamefont {A.}~\bibnamefont {Sacuto}}, \bibinfo {author}
  {\bibfnamefont {D.}~\bibnamefont {Colson}}, \ and\ \bibinfo {author}
  {\bibfnamefont {A.}~\bibnamefont {Forget}},\ }\href@noop {} {\bibfield
  {journal} {\bibinfo  {journal} {Phys. Rev. B}\ }\textbf {\bibinfo {volume}
  {82}},\ \bibinfo {pages} {180521} (\bibinfo {year} {2010})}\BibitemShut
  {NoStop}%
\bibitem [{\citenamefont {Muschler}\ \emph {et~al.}(2009)\citenamefont
  {Muschler}, \citenamefont {Prestel}, \citenamefont {Hackl}, \citenamefont
  {Devereaux}, \citenamefont {Analytis}, \citenamefont {Chu},\ and\
  \citenamefont {Fisher}}]{MuschlerPRB09}%
  \BibitemOpen
  \bibfield  {author} {\bibinfo {author} {\bibfnamefont {B.}~\bibnamefont
  {Muschler}}, \bibinfo {author} {\bibfnamefont {W.}~\bibnamefont {Prestel}},
  \bibinfo {author} {\bibfnamefont {R.}~\bibnamefont {Hackl}}, \bibinfo
  {author} {\bibfnamefont {T.~P.}\ \bibnamefont {Devereaux}}, \bibinfo {author}
  {\bibfnamefont {J.~G.}\ \bibnamefont {Analytis}}, \bibinfo {author}
  {\bibfnamefont {J.-H.}\ \bibnamefont {Chu}}, \ and\ \bibinfo {author}
  {\bibfnamefont {I.~R.}\ \bibnamefont {Fisher}},\ }\href@noop {} {\bibfield
  {journal} {\bibinfo  {journal} {Phys. Rev. B}\ }\textbf {\bibinfo {volume}
  {80}},\ \bibinfo {pages} {180510} (\bibinfo {year} {2009})}\BibitemShut
  {NoStop}%
\bibitem [{\citenamefont {Coldea}\ \emph {et~al.}(2008)\citenamefont {Coldea},
  \citenamefont {Fletcher}, \citenamefont {Carrington}, \citenamefont
  {Analytis}, \citenamefont {Bangura}, \citenamefont {Chu}, \citenamefont
  {Erickson}, \citenamefont {Fisher}, \citenamefont {Hussey},\ and\
  \citenamefont {McDonald}}]{ColdeaPRL08}%
  \BibitemOpen
  \bibfield  {author} {\bibinfo {author} {\bibfnamefont {A.~I.}\ \bibnamefont
  {Coldea}}, \bibinfo {author} {\bibfnamefont {J.~D.}\ \bibnamefont
  {Fletcher}}, \bibinfo {author} {\bibfnamefont {A.}~\bibnamefont
  {Carrington}}, \bibinfo {author} {\bibfnamefont {J.~G.}\ \bibnamefont
  {Analytis}}, \bibinfo {author} {\bibfnamefont {A.~F.}\ \bibnamefont
  {Bangura}}, \bibinfo {author} {\bibfnamefont {J.-H.}\ \bibnamefont {Chu}},
  \bibinfo {author} {\bibfnamefont {A.~S.}\ \bibnamefont {Erickson}}, \bibinfo
  {author} {\bibfnamefont {I.~R.}\ \bibnamefont {Fisher}}, \bibinfo {author}
  {\bibfnamefont {N.~E.}\ \bibnamefont {Hussey}}, \ and\ \bibinfo {author}
  {\bibfnamefont {R.~D.}\ \bibnamefont {McDonald}},\ }\href@noop {} {\bibfield
  {journal} {\bibinfo  {journal} {Phys. Rev. Lett.}\ }\textbf {\bibinfo
  {volume} {101}},\ \bibinfo {pages} {216402} (\bibinfo {year}
  {2008})}\BibitemShut {NoStop}%
\bibitem [{\citenamefont {Shishido}\ \emph {et~al.}(2010)\citenamefont
  {Shishido}, \citenamefont {Bangura}, \citenamefont {Coldea}, \citenamefont
  {Tonegawa}, \citenamefont {Hashimoto}, \citenamefont {Kasahara},
  \citenamefont {Rourke}, \citenamefont {Ikeda}, \citenamefont {Terashima},
  \citenamefont {Settai}, \citenamefont {\ifmmode~\bar{O}\else \={O}\fi{}nuki},
  \citenamefont {Vignolles}, \citenamefont {Proust}, \citenamefont {Vignolle},
  \citenamefont {McCollam}, \citenamefont {Matsuda}, \citenamefont
  {Shibauchi},\ and\ \citenamefont {Carrington}}]{ShishidoPRL10}%
  \BibitemOpen
  \bibfield  {author} {\bibinfo {author} {\bibfnamefont {H.}~\bibnamefont
  {Shishido}}, \bibinfo {author} {\bibfnamefont {A.~F.}\ \bibnamefont
  {Bangura}}, \bibinfo {author} {\bibfnamefont {A.~I.}\ \bibnamefont {Coldea}},
  \bibinfo {author} {\bibfnamefont {S.}~\bibnamefont {Tonegawa}}, \bibinfo
  {author} {\bibfnamefont {K.}~\bibnamefont {Hashimoto}}, \bibinfo {author}
  {\bibfnamefont {S.}~\bibnamefont {Kasahara}}, \bibinfo {author}
  {\bibfnamefont {P.~M.~C.}\ \bibnamefont {Rourke}}, \bibinfo {author}
  {\bibfnamefont {H.}~\bibnamefont {Ikeda}}, \bibinfo {author} {\bibfnamefont
  {T.}~\bibnamefont {Terashima}}, \bibinfo {author} {\bibfnamefont
  {R.}~\bibnamefont {Settai}}, \bibinfo {author} {\bibfnamefont
  {Y.}~\bibnamefont {\ifmmode~\bar{O}\else \={O}\fi{}nuki}}, \bibinfo {author}
  {\bibfnamefont {D.}~\bibnamefont {Vignolles}}, \bibinfo {author}
  {\bibfnamefont {C.}~\bibnamefont {Proust}}, \bibinfo {author} {\bibfnamefont
  {B.}~\bibnamefont {Vignolle}}, \bibinfo {author} {\bibfnamefont
  {A.}~\bibnamefont {McCollam}}, \bibinfo {author} {\bibfnamefont
  {Y.}~\bibnamefont {Matsuda}}, \bibinfo {author} {\bibfnamefont
  {T.}~\bibnamefont {Shibauchi}}, \ and\ \bibinfo {author} {\bibfnamefont
  {A.}~\bibnamefont {Carrington}},\ }\href@noop {} {\bibfield  {journal}
  {\bibinfo  {journal} {Phys. Rev. Lett.}\ }\textbf {\bibinfo {volume} {104}},\
  \bibinfo {pages} {057008} (\bibinfo {year} {2010})}\BibitemShut {NoStop}%
\bibitem [{\citenamefont {Andersen}\ and\ \citenamefont
  {Boeri}(2011)}]{Andersen2011}%
  \BibitemOpen
  \bibfield  {author} {\bibinfo {author} {\bibfnamefont {O.}~\bibnamefont
  {Andersen}}\ and\ \bibinfo {author} {\bibfnamefont {L.}~\bibnamefont
  {Boeri}},\ }\href@noop {} {\bibfield  {journal} {\bibinfo  {journal} {Annalen
  der Physik}\ }\textbf {\bibinfo {volume} {523}},\ \bibinfo {pages} {8}
  (\bibinfo {year} {2011})}\BibitemShut {NoStop}%
\bibitem [{\citenamefont {Graser}\ \emph {et~al.}(2010)\citenamefont {Graser},
  \citenamefont {Kemper}, \citenamefont {Maier}, \citenamefont {Cheng},
  \citenamefont {Hirschfeld},\ and\ \citenamefont {Scalapino}}]{Graser2010}%
  \BibitemOpen
  \bibfield  {author} {\bibinfo {author} {\bibfnamefont {S.}~\bibnamefont
  {Graser}}, \bibinfo {author} {\bibfnamefont {A.~F.}\ \bibnamefont {Kemper}},
  \bibinfo {author} {\bibfnamefont {T.~A.}\ \bibnamefont {Maier}}, \bibinfo
  {author} {\bibfnamefont {H.-P.}\ \bibnamefont {Cheng}}, \bibinfo {author}
  {\bibfnamefont {P.~J.}\ \bibnamefont {Hirschfeld}}, \ and\ \bibinfo {author}
  {\bibfnamefont {D.~J.}\ \bibnamefont {Scalapino}},\ }\href@noop {} {\bibfield
   {journal} {\bibinfo  {journal} {Phys. Rev. B}\ }\textbf {\bibinfo {volume}
  {81}},\ \bibinfo {pages} {214503} (\bibinfo {year} {2010})}\BibitemShut
  {NoStop}%
\bibitem [{\citenamefont {Fuglsang~Jensen}\ \emph {et~al.}(2011)\citenamefont
  {Fuglsang~Jensen}, \citenamefont {Brouet}, \citenamefont {Papalazarou},
  \citenamefont {Nicolaou}, \citenamefont {Taleb-Ibrahimi}, \citenamefont
  {Le~F\`evre}, \citenamefont {Bertran}, \citenamefont {Forget},\ and\
  \citenamefont {Colson}}]{JensenPRB11}%
  \BibitemOpen
  \bibfield  {author} {\bibinfo {author} {\bibfnamefont {M.}~\bibnamefont
  {Fuglsang~Jensen}}, \bibinfo {author} {\bibfnamefont {V.}~\bibnamefont
  {Brouet}}, \bibinfo {author} {\bibfnamefont {E.}~\bibnamefont {Papalazarou}},
  \bibinfo {author} {\bibfnamefont {A.}~\bibnamefont {Nicolaou}}, \bibinfo
  {author} {\bibfnamefont {A.}~\bibnamefont {Taleb-Ibrahimi}}, \bibinfo
  {author} {\bibfnamefont {P.}~\bibnamefont {Le~F\`evre}}, \bibinfo {author}
  {\bibfnamefont {F.}~\bibnamefont {Bertran}}, \bibinfo {author} {\bibfnamefont
  {A.}~\bibnamefont {Forget}}, \ and\ \bibinfo {author} {\bibfnamefont
  {D.}~\bibnamefont {Colson}},\ }\href@noop {} {\bibfield  {journal} {\bibinfo
  {journal} {Phys. Rev. B}\ }\textbf {\bibinfo {volume} {84}},\ \bibinfo
  {pages} {014509} (\bibinfo {year} {2011})}\BibitemShut {NoStop}%
\bibitem [{\citenamefont {Voit}\ \emph {et~al.}(2000)\citenamefont {Voit},
  \citenamefont {Perfetti}, \citenamefont {Zwick}, \citenamefont {Berger},
  \citenamefont {Margaritondo}, \citenamefont {Grüner}, \citenamefont
  {Höchst},\ and\ \citenamefont {Grioni}}]{VoitScience2000}%
  \BibitemOpen
  \bibfield  {author} {\bibinfo {author} {\bibfnamefont {J.}~\bibnamefont
  {Voit}}, \bibinfo {author} {\bibfnamefont {L.}~\bibnamefont {Perfetti}},
  \bibinfo {author} {\bibfnamefont {F.}~\bibnamefont {Zwick}}, \bibinfo
  {author} {\bibfnamefont {H.}~\bibnamefont {Berger}}, \bibinfo {author}
  {\bibfnamefont {G.}~\bibnamefont {Margaritondo}}, \bibinfo {author}
  {\bibfnamefont {G.}~\bibnamefont {Grüner}}, \bibinfo {author} {\bibfnamefont
  {H.}~\bibnamefont {Höchst}}, \ and\ \bibinfo {author} {\bibfnamefont
  {M.}~\bibnamefont {Grioni}},\ }\href@noop {} {\bibfield  {journal} {\bibinfo
  {journal} {Science}\ }\textbf {\bibinfo {volume} {290}},\ \bibinfo {pages}
  {501} (\bibinfo {year} {2000})}\BibitemShut {NoStop}%
\bibitem [{\citenamefont {Brouet}\ \emph {et~al.}(2008)\citenamefont {Brouet},
  \citenamefont {Yang}, \citenamefont {Zhou}, \citenamefont {Hussain},
  \citenamefont {Moore}, \citenamefont {He}, \citenamefont {Lu}, \citenamefont
  {Shen}, \citenamefont {Laverock}, \citenamefont {Dugdale}, \citenamefont
  {Ru},\ and\ \citenamefont {Fisher}}]{Brouet:2008}%
  \BibitemOpen
  \bibfield  {author} {\bibinfo {author} {\bibfnamefont {V.}~\bibnamefont
  {Brouet}}, \bibinfo {author} {\bibfnamefont {W.~L.}\ \bibnamefont {Yang}},
  \bibinfo {author} {\bibfnamefont {X.~J.}\ \bibnamefont {Zhou}}, \bibinfo
  {author} {\bibfnamefont {Z.}~\bibnamefont {Hussain}}, \bibinfo {author}
  {\bibfnamefont {R.~G.}\ \bibnamefont {Moore}}, \bibinfo {author}
  {\bibfnamefont {R.}~\bibnamefont {He}}, \bibinfo {author} {\bibfnamefont
  {D.~H.}\ \bibnamefont {Lu}}, \bibinfo {author} {\bibfnamefont {Z.~X.}\
  \bibnamefont {Shen}}, \bibinfo {author} {\bibfnamefont {J.}~\bibnamefont
  {Laverock}}, \bibinfo {author} {\bibfnamefont {S.~B.}\ \bibnamefont
  {Dugdale}}, \bibinfo {author} {\bibfnamefont {N.}~\bibnamefont {Ru}}, \ and\
  \bibinfo {author} {\bibfnamefont {I.~R.}\ \bibnamefont {Fisher}},\
  }\href@noop {} {\bibfield  {journal} {\bibinfo  {journal} {Phys. Rev. B}\
  }\textbf {\bibinfo {volume} {77}},\ \bibinfo {pages} {235104} (\bibinfo
  {year} {2008})}\BibitemShut {NoStop}%
\bibitem [{\citenamefont {Lin}\ \emph {et~al.}(2011)\citenamefont {Lin},
  \citenamefont {Berlijn}, \citenamefont {Wang}, \citenamefont {Lee},
  \citenamefont {Yin},\ and\ \citenamefont {Ku}}]{LinWeiKu}%
  \BibitemOpen
  \bibfield  {author} {\bibinfo {author} {\bibfnamefont {C.-H.}\ \bibnamefont
  {Lin}}, \bibinfo {author} {\bibfnamefont {T.}~\bibnamefont {Berlijn}},
  \bibinfo {author} {\bibfnamefont {L.}~\bibnamefont {Wang}}, \bibinfo {author}
  {\bibfnamefont {C.-C.}\ \bibnamefont {Lee}}, \bibinfo {author} {\bibfnamefont
  {W.-G.}\ \bibnamefont {Yin}}, \ and\ \bibinfo {author} {\bibfnamefont
  {W.}~\bibnamefont {Ku}},\ }\href@noop {} {\bibfield  {journal} {\bibinfo
  {journal} {Phys. Rev. Lett.}\ }\textbf {\bibinfo {volume} {107}},\ \bibinfo
  {pages} {257001} (\bibinfo {year} {2011})}\BibitemShut {NoStop}%
\bibitem [{\citenamefont {Ortenzi}\ \emph {et~al.}()\citenamefont {Ortenzi},
  \citenamefont {Cappelluti}, \citenamefont {Benfatto},\ and\ \citenamefont
  {Pietronero}}]{OrtenziPRL09}%
  \BibitemOpen
  \bibfield  {author} {\bibinfo {author} {\bibfnamefont {L.}~\bibnamefont
  {Ortenzi}}, \bibinfo {author} {\bibfnamefont {E.}~\bibnamefont {Cappelluti}},
  \bibinfo {author} {\bibfnamefont {L.}~\bibnamefont {Benfatto}}, \ and\
  \bibinfo {author} {\bibfnamefont {L.}~\bibnamefont {Pietronero}},\
  }\href@noop {} {\ }\BibitemShut {NoStop}%
\bibitem [{\citenamefont {Brouet}\ \emph {et~al.}(2009)\citenamefont {Brouet},
  \citenamefont {Marsi}, \citenamefont {Mansart}, \citenamefont {Nicolaou},
  \citenamefont {Taleb-Ibrahimi}, \citenamefont {Le~F\`evre}, \citenamefont
  {Bertran}, \citenamefont {Rullier-Albenque}, \citenamefont {Forget},\ and\
  \citenamefont {Colson}}]{BrouetPRB09}%
  \BibitemOpen
  \bibfield  {author} {\bibinfo {author} {\bibfnamefont {V.}~\bibnamefont
  {Brouet}}, \bibinfo {author} {\bibfnamefont {M.}~\bibnamefont {Marsi}},
  \bibinfo {author} {\bibfnamefont {B.}~\bibnamefont {Mansart}}, \bibinfo
  {author} {\bibfnamefont {A.}~\bibnamefont {Nicolaou}}, \bibinfo {author}
  {\bibfnamefont {A.}~\bibnamefont {Taleb-Ibrahimi}}, \bibinfo {author}
  {\bibfnamefont {P.}~\bibnamefont {Le~F\`evre}}, \bibinfo {author}
  {\bibfnamefont {F.}~\bibnamefont {Bertran}}, \bibinfo {author} {\bibfnamefont
  {F.}~\bibnamefont {Rullier-Albenque}}, \bibinfo {author} {\bibfnamefont
  {A.}~\bibnamefont {Forget}}, \ and\ \bibinfo {author} {\bibfnamefont
  {D.}~\bibnamefont {Colson}},\ }\href@noop {} {\bibfield  {journal} {\bibinfo
  {journal} {Phys. Rev. B}\ }\textbf {\bibinfo {volume} {80}},\ \bibinfo
  {pages} {165115} (\bibinfo {year} {2009})}\BibitemShut {NoStop}%
\bibitem [{\citenamefont {Blaha}\ \emph {et~al.}(2002)\citenamefont {Blaha}
  \emph {et~al.}}]{Wien2k}%
  \BibitemOpen
  \bibfield  {author} {\bibinfo {author} {\bibfnamefont {P.}~\bibnamefont
  {Blaha}} \emph {et~al.},\ }\href@noop {} {\bibfield  {journal} {\bibinfo
  {journal} {Wien2K, ISBN 3-9501031-1-2}\ } (\bibinfo {year}
  {2002})}\BibitemShut {NoStop}%
\bibitem [{\citenamefont {Lee}\ \emph {et~al.}(2009)\citenamefont {Lee},
  \citenamefont {Yin},\ and\ \citenamefont {Ku}}]{Lee:2009}%
  \BibitemOpen
  \bibfield  {author} {\bibinfo {author} {\bibfnamefont {C.-C.}\ \bibnamefont
  {Lee}}, \bibinfo {author} {\bibfnamefont {W.-G.}\ \bibnamefont {Yin}}, \ and\
  \bibinfo {author} {\bibfnamefont {W.}~\bibnamefont {Ku}},\ }\href@noop {}
  {\bibfield  {journal} {\bibinfo  {journal} {Phys. Rev. Lett.}\ }\textbf
  {\bibinfo {volume} {103}},\ \bibinfo {pages} {267001} (\bibinfo {year}
  {2009})}\BibitemShut {NoStop}%
\bibitem [{\citenamefont {Ku}\ \emph {et~al.}(2010)\citenamefont {Ku},
  \citenamefont {Berlijn},\ and\ \citenamefont {Lee}}]{WeiKuPRL2010}%
  \BibitemOpen
  \bibfield  {author} {\bibinfo {author} {\bibfnamefont {W.}~\bibnamefont
  {Ku}}, \bibinfo {author} {\bibfnamefont {T.}~\bibnamefont {Berlijn}}, \ and\
  \bibinfo {author} {\bibfnamefont {C.-C.}\ \bibnamefont {Lee}},\ }\href@noop
  {} {\bibfield  {journal} {\bibinfo  {journal} {Phys. Rev. Lett.}\ }\textbf
  {\bibinfo {volume} {104}},\ \bibinfo {pages} {216401} (\bibinfo {year}
  {2010})}\BibitemShut {NoStop}%
\bibitem [{\citenamefont {Mansart}\ \emph {et~al.}(2011)\citenamefont
  {Mansart}, \citenamefont {Brouet}, \citenamefont {Papalazarou}, \citenamefont
  {Fuglsang~Jensen}, \citenamefont {Petaccia}, \citenamefont {Gorovikov},
  \citenamefont {Grum-Grzhimailo}, \citenamefont {Rullier-Albenque},
  \citenamefont {Forget}, \citenamefont {Colson},\ and\ \citenamefont
  {Marsi}}]{MansartPRB11}%
  \BibitemOpen
  \bibfield  {author} {\bibinfo {author} {\bibfnamefont {B.}~\bibnamefont
  {Mansart}}, \bibinfo {author} {\bibfnamefont {V.}~\bibnamefont {Brouet}},
  \bibinfo {author} {\bibfnamefont {E.}~\bibnamefont {Papalazarou}}, \bibinfo
  {author} {\bibfnamefont {M.}~\bibnamefont {Fuglsang~Jensen}}, \bibinfo
  {author} {\bibfnamefont {L.}~\bibnamefont {Petaccia}}, \bibinfo {author}
  {\bibfnamefont {S.}~\bibnamefont {Gorovikov}}, \bibinfo {author}
  {\bibfnamefont {A.~N.}\ \bibnamefont {Grum-Grzhimailo}}, \bibinfo {author}
  {\bibfnamefont {F.}~\bibnamefont {Rullier-Albenque}}, \bibinfo {author}
  {\bibfnamefont {A.}~\bibnamefont {Forget}}, \bibinfo {author} {\bibfnamefont
  {D.}~\bibnamefont {Colson}}, \ and\ \bibinfo {author} {\bibfnamefont
  {M.}~\bibnamefont {Marsi}},\ }\href@noop {} {\bibfield  {journal} {\bibinfo
  {journal} {Phys. Rev. B}\ }\textbf {\bibinfo {volume} {83}},\ \bibinfo
  {pages} {064516} (\bibinfo {year} {2011})}\BibitemShut {NoStop}%
\bibitem [{\citenamefont {A.~Damascelli}\ and\ \citenamefont
  {Shen}(2003)}]{DamascelliRMP2003}%
  \BibitemOpen
  \bibfield  {author} {\bibinfo {author} {\bibfnamefont {Z.~H.}\ \bibnamefont
  {A.~Damascelli}}\ and\ \bibinfo {author} {\bibfnamefont {Z.-X.}\ \bibnamefont
  {Shen}},\ }\href@noop {} {\bibfield  {journal} {\bibinfo  {journal} {Rev.
  Mod. Phys.}\ }\textbf {\bibinfo {volume} {75}},\ \bibinfo {pages} {473}
  (\bibinfo {year} {2003})}\BibitemShut {NoStop}%
\bibitem [{\citenamefont {Asensio}\ \emph {et~al.}(2003)\citenamefont
  {Asensio}, \citenamefont {Avila}, \citenamefont {Roca}, \citenamefont
  {Tejeda}, \citenamefont {Gu}, \citenamefont {Lindroos}, \citenamefont
  {Markiewicz},\ and\ \citenamefont {Bansil}}]{Asensio:2003}%
  \BibitemOpen
  \bibfield  {author} {\bibinfo {author} {\bibfnamefont {M.~C.}\ \bibnamefont
  {Asensio}}, \bibinfo {author} {\bibfnamefont {J.}~\bibnamefont {Avila}},
  \bibinfo {author} {\bibfnamefont {L.}~\bibnamefont {Roca}}, \bibinfo {author}
  {\bibfnamefont {A.}~\bibnamefont {Tejeda}}, \bibinfo {author} {\bibfnamefont
  {G.~D.}\ \bibnamefont {Gu}}, \bibinfo {author} {\bibfnamefont
  {M.}~\bibnamefont {Lindroos}}, \bibinfo {author} {\bibfnamefont {R.~S.}\
  \bibnamefont {Markiewicz}}, \ and\ \bibinfo {author} {\bibfnamefont
  {A.}~\bibnamefont {Bansil}},\ }\href@noop {} {\bibfield  {journal} {\bibinfo
  {journal} {Phys. Rev. B}\ }\textbf {\bibinfo {volume} {67}},\ \bibinfo
  {pages} {014519} (\bibinfo {year} {2003})}\BibitemShut {NoStop}%
\bibitem [{\citenamefont {Wang}\ \emph {et~al.}()\citenamefont {Wang},
  \citenamefont {Richard}, \citenamefont {Huang}, \citenamefont {Miao},
  \citenamefont {Cevey}, \citenamefont {Xu}, \citenamefont {Sun}, \citenamefont
  {Qian}, \citenamefont {Xu}, \citenamefont {Shi}, \citenamefont {Hu},
  \citenamefont {Dai},\ and\ \citenamefont {Ding}}]{Hong2012}%
  \BibitemOpen
  \bibfield  {author} {\bibinfo {author} {\bibfnamefont {X.}~\bibnamefont
  {Wang}}, \bibinfo {author} {\bibfnamefont {P.}~\bibnamefont {Richard}},
  \bibinfo {author} {\bibfnamefont {Y.-B.}\ \bibnamefont {Huang}}, \bibinfo
  {author} {\bibfnamefont {H.}~\bibnamefont {Miao}}, \bibinfo {author}
  {\bibfnamefont {L.}~\bibnamefont {Cevey}}, \bibinfo {author} {\bibfnamefont
  {N.}~\bibnamefont {Xu}}, \bibinfo {author} {\bibfnamefont {Y.-J.}\
  \bibnamefont {Sun}}, \bibinfo {author} {\bibfnamefont {T.}~\bibnamefont
  {Qian}}, \bibinfo {author} {\bibfnamefont {Y.-M.}\ \bibnamefont {Xu}},
  \bibinfo {author} {\bibfnamefont {M.}~\bibnamefont {Shi}}, \bibinfo {author}
  {\bibfnamefont {J.-P.}\ \bibnamefont {Hu}}, \bibinfo {author} {\bibfnamefont
  {X.}~\bibnamefont {Dai}}, \ and\ \bibinfo {author} {\bibfnamefont
  {H.}~\bibnamefont {Ding}},\ }\href@noop {} {\bibinfo  {journal}
  {arXiv:1201.3655 (2012).}\ }\BibitemShut {NoStop}%
\bibitem [{\citenamefont {Park}\ \emph {et~al.}(2010)\citenamefont {Park},
  \citenamefont {Inosov}, \citenamefont {Yaresko}, \citenamefont {Graser},
  \citenamefont {Sun}, \citenamefont {Bourges}, \citenamefont {Sidis},
  \citenamefont {Li}, \citenamefont {Kim}, \citenamefont {Haug}, \citenamefont
  {Ivanov}, \citenamefont {Hradil}, \citenamefont {Schneidewind}, \citenamefont
  {Link}, \citenamefont {Faulhaber}, \citenamefont {Glavatskyy}, \citenamefont
  {Lin}, \citenamefont {Keimer},\ and\ \citenamefont {Hinkov}}]{ParkPRB10}%
  \BibitemOpen
\bibfield  {journal} {  }\bibfield  {author} {\bibinfo {author} {\bibfnamefont
  {J.~T.}\ \bibnamefont {Park}}, \bibinfo {author} {\bibfnamefont {D.~S.}\
  \bibnamefont {Inosov}}, \bibinfo {author} {\bibfnamefont {A.}~\bibnamefont
  {Yaresko}}, \bibinfo {author} {\bibfnamefont {S.}~\bibnamefont {Graser}},
  \bibinfo {author} {\bibfnamefont {D.~L.}\ \bibnamefont {Sun}}, \bibinfo
  {author} {\bibfnamefont {P.}~\bibnamefont {Bourges}}, \bibinfo {author}
  {\bibfnamefont {Y.}~\bibnamefont {Sidis}}, \bibinfo {author} {\bibfnamefont
  {Y.}~\bibnamefont {Li}}, \bibinfo {author} {\bibfnamefont {J.-H.}\
  \bibnamefont {Kim}}, \bibinfo {author} {\bibfnamefont {D.}~\bibnamefont
  {Haug}}, \bibinfo {author} {\bibfnamefont {A.}~\bibnamefont {Ivanov}},
  \bibinfo {author} {\bibfnamefont {K.}~\bibnamefont {Hradil}}, \bibinfo
  {author} {\bibfnamefont {A.}~\bibnamefont {Schneidewind}}, \bibinfo {author}
  {\bibfnamefont {P.}~\bibnamefont {Link}}, \bibinfo {author} {\bibfnamefont
  {E.}~\bibnamefont {Faulhaber}}, \bibinfo {author} {\bibfnamefont
  {I.}~\bibnamefont {Glavatskyy}}, \bibinfo {author} {\bibfnamefont {C.~T.}\
  \bibnamefont {Lin}}, \bibinfo {author} {\bibfnamefont {B.}~\bibnamefont
  {Keimer}}, \ and\ \bibinfo {author} {\bibfnamefont {V.}~\bibnamefont
  {Hinkov}},\ }\href@noop {} {\bibfield  {journal} {\bibinfo  {journal} {Phys.
  Rev. B}\ }\textbf {\bibinfo {volume} {82}},\ \bibinfo {pages} {134503}
  (\bibinfo {year} {2010})}\BibitemShut {NoStop}%
\end{thebibliography}%

\end{document}